\documentclass[twocolumn, aps, pra, superscriptaddress, 10pt]{revtex4-2}
\usepackage{gensymb}
\usepackage{graphicx}
\usepackage{amsmath, amssymb}
\usepackage{physics}
\usepackage{caption}
\usepackage{subcaption}
\usepackage{xcolor}
\usepackage{svg}
\usepackage{cases}
\usepackage{tabularx}
\usepackage{tikz}
\usepackage{tikz-3dplot}
\usepackage{overpic}
\usetikzlibrary{angles, arrows.meta, calc, quotes}
\usepackage{amsmath}
\usetikzlibrary{positioning}
\usepackage{ragged2e}
\usetikzlibrary{decorations.pathreplacing}
\tdplotsetmaincoords{70}{110}  
\usepackage{hyperref}
\begin{document}


    \title{Optical signatures of coherence in molecular dimers}
    
    \author{Priyankar Banerjee}
    \email{pb2049@hw.ac.uk}
    \affiliation{SUPA, Institute of Photonics and Quantum Sciences, Heriot-Watt University, Edinburgh EH14 4AS, Scotland, United Kingdom}
    \author{Adam Burgess}
    \affiliation{SUPA, Institute of Photonics and Quantum Sciences, Heriot-Watt University, Edinburgh EH14 4AS, Scotland, United Kingdom}
    \author{Julian Wiercinski}
    \affiliation{SUPA, Institute of Photonics and Quantum Sciences, Heriot-Watt University, Edinburgh EH14 4AS, Scotland, United Kingdom}
    \author{Moritz Cygorek}
    \affiliation{Condensed Matter Theory, Department of Physics, TU Dortmund, 44221 Dortmund, Germany}
    \author{Erik M. Gauger}
    \affiliation{SUPA, Institute of Photonics and Quantum Sciences, Heriot-Watt University, Edinburgh EH14 4AS, Scotland, United Kingdom}

    \begin{abstract}
        We calculate experimentally measurable signatures of quantum correlations in a coupled molecular dimer that strongly interacts with its vibrational environment. 
        We investigate intensity and mode-resolved photon coincidences for different relative orientations of such dimers, and observe spatio-temporal correlations for various configurations.
        We find that projective measurements can produce cooperative signatures even when emitters are arranged orthogonal to each other.
        To model effects of vibrational environments that are present in realistic experimental situations, we use the polaron framework. Further, we also account for the effects of finite instrument response, varying temperature, and presence of static disorder.
        We analyse the effect of disorder in both dimer orientation and measurement direction and find that photon coincidences remain well-resolvable using state-of-the-art detectors.
        This work enhances our understanding of cooperative emission from two coupled emitters and offers direction for future experiments on probing their coherent dynamics.
    \end{abstract}

    \maketitle


    \section{Introduction}
        \label{sec:intro} 
        Quantum coherence can play a pivotal role in energy transport, quantum sensing, and light-harvesting processes \cite{Lambert2013, Collini2010, doi:10.1073/pnas.1005484107, 10.1063/1.3002335}.  
        Experimental detection of quantum correlations in solid-state platforms \cite{PhysRevLett.118.233602, stotz2005coherent, tran2016quantum} and bio-molecular processes \cite{Mattioni2024, 10.1063/1.5110275, Polisseni:16, scholes2017using, Huelga01072013, scholes2011lessons} is a major scientific challenge, crucial for the development of next generation quantum technologies \cite{toninelli2021single,PhysRevResearch.5.043097,PhysRevA.82.062310}.
        Characterising quantum effects can be particularly challenging in the presence of deleterious effects like molecular vibrations and presence of static disorder.  
        However, phonon interactions have also been proposed to play a beneficial role in coherent exciton transfer \cite{plenio2008dephasing, rebentrost2009environment,Tomasi2020} in molecular complexes, efficient transport along molecular chains \cite{PRXQuantum.3.020354,burgess2025enhancing}, optimal power generation \cite{Rouse_2019, PhysRevLett.111.253601, C4CP05310A} and ``optical ratcheting" \cite{PRXEnergy.2.013002} in bio-inspired configurations of nano-emitters. 
        Understanding how phonons influence these processes is therefore crucial for designing robust quantum mechanical light-harvesting devices \cite{PhysRevLett.111.253601, PhysRevLett.117.203603, C4CP05310A, doi:10.1073/pnas.1212666110, PhysRevLett.104.207701, doi:10.1021/acs.jpcc.7b07138, C8CP05535A, C6CP06098F}. 
        The role of long-lived coherences in photosynthetic complexes remains debated, with evidence suggesting many observed signals stem from vibrational rather than electronic coherence and decay too rapidly to influence energy transfer \cite{doi:10.1126/sciadv.aaz4888, doi:10.1073/pnas.1702261114, thyrhaug2018identification}. 
        Here, we focus on well-established inter-emitter (excitonic) coherences, manifesting as delocalised eigenstates in coupled two-level systems. 
        Throughout this work, ``coherence'' specifically refers to off-diagonal density-matrix elements in the eigenbasis of the system Hamiltonian.
        
        Molecular aggregates, the building blocks of light-harvesting systems, often dimerize to form two distinct hybridised states with unique optical properties \cite{sarovar2010quantum, doi:10.1021/jp963777g, https://doi.org/10.1111/j.1751-1097.1985.tb01655.x}.
        By tuning the dipole strength, relative position and orientation of optical dipoles, one can control the energy splitting of the single exciton eigenstates \cite{doi:10.1021/acs.chemrev.7b00581}.
        This results in the formation of distinct excitonic states, commonly referred to as bright and dark states. 
        The bright state, due to an enhanced transition dipole moment, exhibits strong optical coupling and can efficiently absorb and transfer energy \cite{may2023charge}. 
        In contrast to the optically active bright states, the dark state lacks significant dipole moment but can be populated through non-radiative phonon-assisted transitions from the bright state \cite{Rouse_2019, PhysRevLett.111.253601}.
        The interplay between the bright and dark states plays a crucial role in the emission properties of molecular dimers \cite{doi:10.1021/acs.chemrev.7b00581}.
        
        Most contemporary methods for experimentally investigating quantum coherence in molecular emitters, such as ultrafast multidimensional spectroscopy, rely on non-linear techniques to probe electronic dynamics \cite{doi:10.1021/jp7107889, engel2007evidence, sung2015direct}. 
        Among these, methods like pump-probe \cite{SAVIKHIN1997303, C9FD00068B}, photon-echo spectroscopy \cite{mukamel1995principles} and 2D spectroscopy \cite{doi:10.1021/jp7107889, engel2007evidence, halpin2014two, duan2015origin, tiwari2013electronic} have shown promise in probing coherent signatures in light-harvesting systems.
        Recent works \cite{PhysRevLett.131.143601,li2023single,PhysRevB.95.201305,PhysRevLett.124.203601,PhysRevA.107.023718,PhysRevA.98.063828} have also illustrated how photon coincidence experiments can serve as powerful tools for probing and characterizing quantum coherence across a diverse range of quantum emitters.
        One such experiment is the Hanbury-Brown-Twiss (HBT) setup \cite{doi:10.1021/acs.nanolett.8b01133, prasad2020correlating}, where emitted photons are split by a beam splitter and directed to two independent detectors, enabling the measurement of temporal correlations between detection events. 
        The resulting second-order photon correlation, provides insight into the quantum statistical nature of the light \cite{PhysRev.130.2529, Grangier1986}.
        Importantly, these photon correlations are not solely properties of the light itself, but also encode information about the quantum state of the emitters \cite{michler2000quantum, 10.1063/1.1415346} and its value is indicative of the presence of interemitter coherence, as the underpinning characteristic cooperative emission \cite{PhysRevResearch.5.013176, PhysRevResearch.3.033136}. 
        Beyond their use in characterising quantum correlations, photon coincidence measurements exhibit sensitivity to spatial effects arising from the system configuration and direction of photon detection. 
        In particular, recent studies with ultra-cold atoms \cite{PhysRevLett.124.063603, PhysRevResearch.5.013163} and solid-state quantum emitters \cite{PhysRevA.107.023718} have shown that the direction along which a photon is sampled can introduce correlations between the quantum emitters. 
        
        In this article, we develop a theory for calculating optical signatures of quantum correlations in a molecular dimer.
        We calculate emission intensity and intensity correlations by sampling photons along different detection directions.
        These observables reveal coherent features in the emission --- such as interference patterns or directional dependencies --- that originate from quantum superpositions of states within the dimer. 
        We refer to such features as coherent signatures, as they serve as indirect evidence of underlying quantum coherence between the emitters.
        Molecular dimers are typically strongly coupled to their vibrational environments \cite{Rouse_2019, Mathew2014, https://doi.org/10.1002/adma.201605497, 10.1063/1.5049537}, which can be accounted for by moving into the polaron picture \cite{mahan2013many, 10.1063/1.2977974, McCutcheon_2010, Nazir_2016}.
        The polaron transformation has been used to study the role of strong phonon coupling in the emission characteristics of a pair of coupled and uncoupled quantum emitters \cite{Rouse_2019, PhysRevResearch.6.033231}, but not in the context of intensity correlation for a coupled molecular dimer. 
        We explore different dimer geometries and find certain dipole orientation where projective measurements along specific directions play a key role in determining the excitation pathway \cite{PhysRevA.107.023718}. 
        Meanwhile, strong phonon coupling and ensemble averaging also determine the resolvability of coherent signatures.
        We find that molecular dimers are surprisingly robust against disorder, and detection of these signatures is limited by finite instrument resolution.
        
        This paper is organised as follows: In Sec.~\ref{sec:model}, we layout the model of a F\"orster-coupled molecular dimer and incorporate strong vibrational coupling by performing the polaron transformation.
        Sec.~\ref{sec:sig_sup} looks into the signatures of interemitter correlations \cite{10.1063/1.460185, 10.1063/1.457174}, analysing photon intensities and two-photon coincidences considering mode-selective detectors.
        We further investigate how projective measurements can show cooperative signatures for orthogonal dimers and induce coherent oscillations in an intermediate $45\degree$ dimer.
        In Sec.~\ref{sec:factors}, we explore the effects of the phonon bath temperature and ensemble averaging and find how they affect the coherent signatures. 
        We then move beyond static dipoles, by taking into account disorder present in realistic experiments.  We average over both the relative orientation of the dimer and the detection angle, examining their impact on photon correlations. 
        Finally, we summarize our findings in Sec.~\ref{sec:conc}.


    \section{Model}
        \label{sec:model}
\begin{figure}[h]
        \centering
        \begin{overpic}[width=0.45\textwidth]{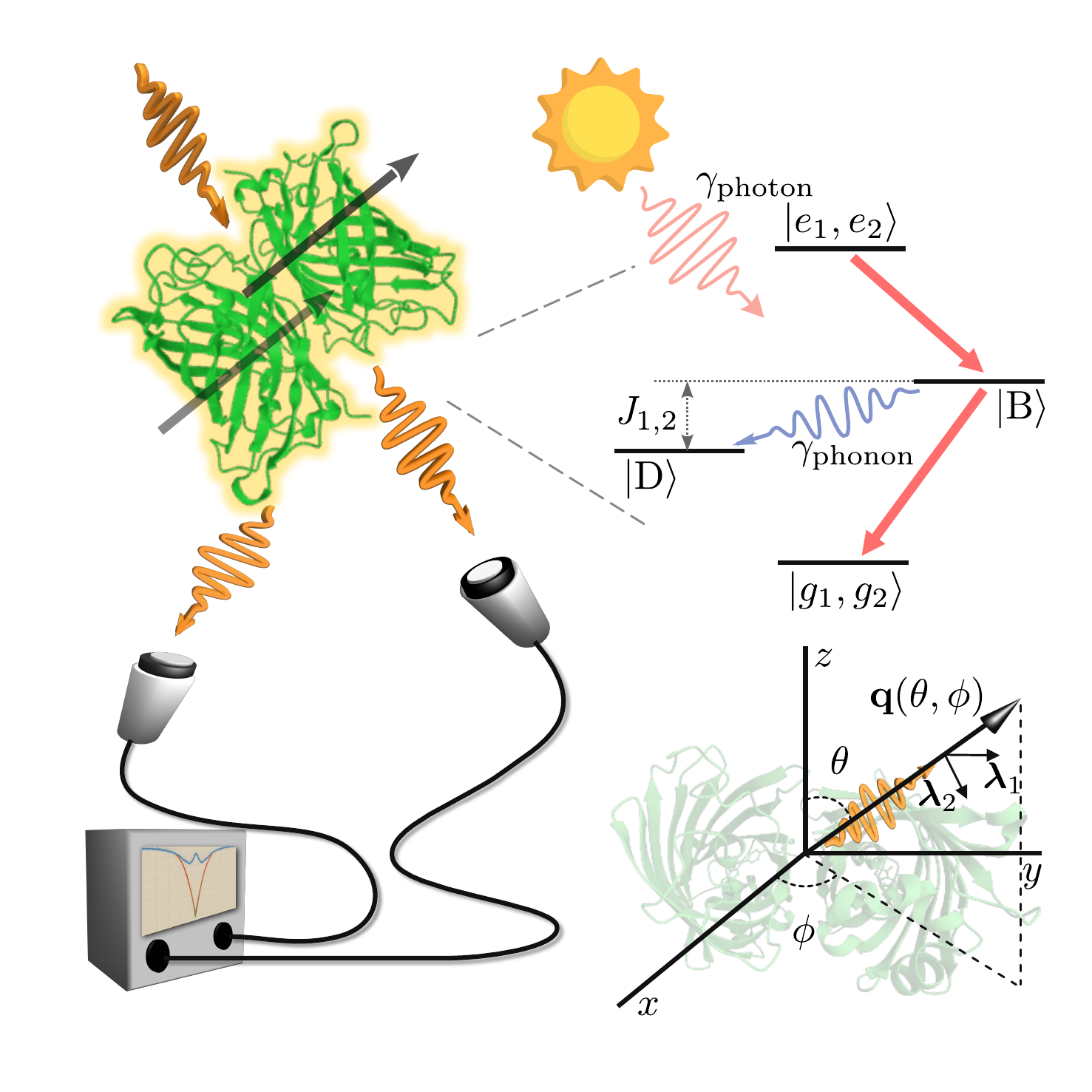}
        \put(-5,88){
                \begin{tikzpicture}
                \node[anchor=north west] at (0, 100) {(a)};
                \end{tikzpicture}}
        \put(53,35){
                \begin{tikzpicture}
                \node[anchor=north west] at (0, 100) {(b)};
                \end{tikzpicture}}
        \end{overpic}
        \caption{\justifying (a) Sketch of a photon coincidence measurement setup. The zoomed-in image shows a level scheme for a H-dimer (dipoles aligned parallel to each other resulting in a positive dipole interaction).
        The rate of incoherent sunlight pumping is given as $\gamma_{\text{photon}}$ and the rate of leakage into the dark state as $\gamma_{\text{phonon}}$. 
        The bright and dark states are shown as $|B \rangle$ and $| D \rangle$ respectively. 
        (b) This panel illustrates a photon detection direction which is associated with a certain mode $\boldsymbol{q}(\theta, \phi)$.
        Each such mode $\boldsymbol{q}$ of an emitted photon is associated with two polarisation directions $\boldsymbol{\lambda}_{1}$ and $\boldsymbol{\lambda}_{2}$.}
            \label{fig:system}
        \end{figure}

        We consider a dimer composed of two monomers with the same optical dipole moment interacting with one another and with their surrounding environments as shown in Fig.~\ref{fig:system}(a).
        These dipoles can be approximated as two-level systems with ground and excited states $\ket{g_{m}}$ and $\ket{e_{m}}$, respectively, where $m \in {1,2}$. Thus, using natural units $(\hbar = c \equiv 1)$, the Hamiltonian describing this system is,
        \begin{equation} 
            \label{eqn:system_ham}
            H_{S} = \omega_{S} \sum_{m=1}^{2} \sigma_{m}^{+} \sigma_{m}^{-} + \frac{J_{1,2}(\boldsymbol{r}_{1,2})}{2} (\sigma_{1}^{+} \sigma_{2}^{-} + \sigma_{1}^{-} \sigma_{2}^{+}),
        \end{equation}  
        where $\omega_{S}$ is the transition frequency of each dipole whose magnitude is assumed to be $1.8$ eV \cite{Pachón_2017, doi:10.1073/pnas.1402538111,B704962E}. 
        Here, we have introduced the raising and lowering operator for the $m$\textsuperscript{th} dipole as $\sigma_{m}^{+} = \ket{e_{m}}\bra{g_{m}}$ and $\sigma_{m}^{-} = \ket{g_{m}}\bra{e_{m}}$, respectively.
        The first term in the Hamiltonian accounts for the energy of the individual dipoles, while the second term accounts for the interaction between them. 
        The latter describes the resonant F\"orster-type interaction between the dipoles \cite{PhysRevA.62.013413} and is given as
        \begin{equation}
            \label{eqn:forst_coup}
            J_{1,2}(\boldsymbol{r}_{1,2}) = \frac{1}{4 \pi \epsilon_{0}} \left(\frac{\boldsymbol{\mu}_{1} \cdot \boldsymbol{\mu}_{2}}{|\boldsymbol{r}^{3}_{1,2}|} - \frac{3(\boldsymbol{r}_{1,2} \cdot \boldsymbol{\mu}_{1})(\boldsymbol{r}_{1,2} \cdot \boldsymbol{\mu}_{2})}{|\boldsymbol{r}^{5}_{1,2}|}\right).
        \end{equation}
        Here, the dipole-dipole coupling strength depends on the relative separation $\boldsymbol{r}_{1,2} = |\boldsymbol{r}_{1} - \boldsymbol{r}_{2}|$ and the orientation of the optical transition dipoles $\boldsymbol{\mu}_{1}$ and $\boldsymbol{\mu}_{2}$. 
        Throughout this paper, we consider the two emitters as being located at positions $\boldsymbol{r}_{1} = - \boldsymbol{r} / 2$ and $\boldsymbol{r}_{2} = \boldsymbol{r} / 2$. 
        We assume this separation to be $\boldsymbol{r} = 2 \boldsymbol{\hat{z}}$ nm. 
        The magnitude of the optical dipole moment is $|\boldsymbol{\mu}_{m}| = 10$ Debye that is typical of chromophores \cite{Pachón_2017, doi:10.1073/pnas.1402538111,B704962E}.
    
        The dimer system interacts with a common multimode optical environment, that is associated with the electromagnetic field
        \begin{equation}\label{eq:em_field}
            \boldsymbol{E}(\boldsymbol{r}_{m}) = i \sum_{\boldsymbol{q}, \lambda} 
            \sqrt{\frac{\omega_{\boldsymbol{q}}}{2\epsilon_0 \hbar \mathcal{V}}}\boldsymbol{e}_{\boldsymbol{q}, \lambda} (a_{\boldsymbol{q}, \lambda} e^{i \boldsymbol{q} \cdot \boldsymbol{r}_{m}}- a_{\boldsymbol{q}, \lambda}^{\dagger}e^{-i \boldsymbol{q} \cdot \boldsymbol{r}_{m}})
        \end{equation}
        with free space permittivity $\epsilon_{0}$ and normalisation volume $\mathcal{V}$. 
        The vectors $\boldsymbol{e}_{\boldsymbol{q}, \lambda}$ describe the polarization directions. The operators $a_{\boldsymbol{q}, \lambda}$ and $a_{\boldsymbol{q}, \lambda}^{\dagger}$ correspond to the annihilation and creation of photons with wave-vector $\boldsymbol{q}$ and polarization $\lambda$, respectively.
        Under the dipole approximation, the interaction of the $m^{\text{th}}$ optical dipole with the optical bath is written as $-\boldsymbol{\mu}_{m} \cdot \boldsymbol{E}$. 
        Thus, the optical interaction takes the form,
        \begin{equation} 
            \label{eqn:H_opt}
            H_{\text{I,opt}} = \sum_{\boldsymbol{q}, \lambda} \sum_{m} \boldsymbol{\mu}_{m} \cdot \boldsymbol{u}_{\boldsymbol{q}, \lambda}(\boldsymbol{r}_{m}) \sigma_{m}^{x}  a_{\boldsymbol{q}, \lambda} + \text{H.c.,}
        \end{equation}
        where $\boldsymbol{u}_{\boldsymbol{q}, \lambda}(\boldsymbol{r}_{m}) = i\sqrt{\omega_{\boldsymbol{q}} / (2\epsilon_0 \hbar \mathcal{V})}\boldsymbol{e}_{\boldsymbol{q}, \lambda}e^{i\boldsymbol{q}\cdot \boldsymbol{r}_{m}}$ represents the spatial mode functions of the light field at the position of the $m\textsuperscript{th}$ dipole and H.c. is the Hermitian conjugate. 
        
        The monomers couple to their local vibrational baths with a linear interaction, represented by displacements of the excited states and can be written as \cite{Rouse_2019},
        \begin{equation} 
            \label{eqn:H_vib}
            H_{\text{I,vib}} = \sum_{m=1}^{2}(\sigma_{m}^{+} \sigma_{m}^{-}) \sum_{\boldsymbol{k}} g_{\boldsymbol{k}}(b_{m,\boldsymbol{k}} + b_{m,\boldsymbol{k}}^{\dagger}).
        \end{equation}
        Here, $g_{\boldsymbol{k}}$ is the coupling strength and $b_{m,\boldsymbol{k}}^{(\dagger)}$ are the annihilation (creation) operators for phonons with wave-vector $\boldsymbol{k}$ of the $m\textsuperscript{th}$ monomer, respectively.
        
        Finally, the free evolution of the optical and phonon environment is modelled using 
        \begin{equation}
            \begin{aligned}
                H_{B} &= \sum_{\boldsymbol{q}, \lambda} \omega_{\boldsymbol{q}, \lambda} a^{\dagger}_{\boldsymbol{q}, \lambda}a_{\boldsymbol{q}, \lambda} + \sum_{m,\boldsymbol{k}} \omega_{m,\boldsymbol{k}} b^{\dagger}_{m,\boldsymbol{k}} b_{m,\boldsymbol{k}}, 
            \end{aligned}
        \end{equation}
        where the first term represents the energy of the optical bath modes, while the second term represents the energy of the vibrational bath modes. 
            
        Typically for molecular systems, the strong coupling to the vibrational environment is tackled by performing a polaron transformation \cite{PhysRevLett.117.203603, PhysRevB.84.081305, Nazir_2016}, which effectively diagonalises part of the system-environment interaction Hamiltonian, allowing the remaining interactions to be treated perturbatively within a weak-coupling framework. 
        In the polaron frame, the phonon environment and its coupling to the system are modified, while the photon environment remains unchanged.

        The Born-Markov master equation captures the system dynamics by tracing out the environmental degrees of freedom and, in the interaction picture, takes the form \cite{10.1093/acprof:oso/9780199213900.001.0001},
        \begin{equation}
        \label{eqn:born_markov}
            \frac{d }{dt}\rho_{S}^{\prime}(t) = - \int^{\infty}_{0} \text{Tr}_{B}\left[H_{I}^{\prime}(t),\left[H_{I}^{\prime}(t-s), \rho_{S}^{\prime}(t) \otimes \rho_{B}^{\prime} \right]\right]ds,
        \end{equation} 
        where $H_{I}^{\prime}(t)$ is the interaction Hamiltonian, and $\rho_B^{\prime}$ is the bath density matrix.
        Here, we have used a prime to denote operators in the polaron frame.
        The Born approximation, central to Eq.~\eqref{eqn:born_markov}, assumes that the environment remains nearly unaffected by the system, while the Markov approximation assumes a short environmental memory time \cite{10.1093/acprof:oso/9780199213900.001.0001}.
        In the polaron frame, these approximations apply to the residual interactions after the transformation, which are assumed to be sufficiently weak and rapidly decaying, even though the original system-phonon coupling may be strong.

        Thus, the non-secular Bloch-Redfield master equation in the polaron frame (see Appendix~\ref{sec:polaron_brme}) is given by \cite{doi:10.1021/acs.jpclett.9b01349, PRXEnergy.2.013002, Rouse_2019}
            \begin{equation} \label{eq:polaron_me}
               \frac{d}{d t} \rho_{S}^{\prime} = -i \left [\hat{H}_{S}^{\prime}, \rho_{S}^{\prime} \right] + \mathcal{D}_{\text{opt}}^{\prime} (\rho_{S}^{\prime})  +  \mathcal{D}_{\text{coup}}^{\prime} (\rho_{S}^{\prime}).
            \end{equation}
        The first term on the right-hand side in Eq.~\eqref{eq:polaron_me} captures the coherent evolution under the influence of the system Hamiltonian $H_{S}^{\prime}$.
        Transforming into the polaron frame rescales the transition energies and coupling in the dimer Hamiltonian, where the frequency is shifted by the reorganisation energy of the phonon environment $\lambda_{0}$ \cite{C7CP06237K, doi:10.1021/acs.jpclett.9b01349, PRXEnergy.2.013002, Rouse_2019}. 
        The second and the third terms capture the dissipative effects due to the rescaled interaction with the surrounding environment and is obtained by a second-order perturbation in the system-bath couplings. 
        The second term $\mathcal{D}_{\text{opt}}^{\prime}$ captures the dissipative dynamics due to a rescaled optical interaction.
        The F\"orster coupling term in Eq.~\eqref{eqn:forst_coup} also introduces an additional interaction term in the polaron frame whose effect is captured by $\mathcal{D}_{\text{coup}}^{\prime}$, which is second order in dipole-dipole coupling.
        For the remainder of this paper, we assume a super-Ohmic spectral density to model the vibrational environment $\mathcal{J}_{\textrm{vib}}(\omega) = \lambda_{0}\omega^{3} \exp[-\omega /\omega_c] / 2 \omega_{c}^{3}$.
        We choose the phonon reorganisation energy $\lambda_{0} = 5$ meV and the cut-off frequency of the bath $\omega_{c} = 90$ meV as in \cite{C7CP06237K, Rouse_2019}.
        This choice of \(\omega_c\) is consistent with estimates for certain photosynthetic systems~\cite{doi:10.1021/acs.jpclett.1c01303, Pachón_2017, B704962E, singh2021coherent}, 
        and its being significantly larger than the dipole-dipole coupling \(J_{1,2}\) justifies the use of the polaron-transformed Bloch–Redfield formalism~\cite{C7CP06237K, Rouse_2019, pollock2013multi}.
        We also set the temperature of the optical bath at $5800$ K, i.e. the solar temperature \cite{C6CP06098F, PhysRevLett.117.203603, Wrfel2005, doi:10.1021/acs.jpcc.7b07138} and the vibrational bath at 300 K.


    \section{Signatures of interemitter Coherence}
    \label{sec:sig_sup}
        The degeneracy in the single excitation eigenstates of molecular dimers is lifted by the presence of dipole-dipole coupling. 
        The relative orientation of dipoles determines the nature of the dipole interaction in the Hamiltonian [Eq.~\eqref{eqn:forst_coup}], which in turn determines the energy ordering of the bright and dark states  \cite{spano2010spectral, knoester2006modeling}.
        In this paper, we investigate collective excitonic behaviour in H- and J-dimers, in addition to intermediate dimer configurations.
        The H- and J-dimers exhibit positive and negative dipole couplings, respectively, due to their characteristic dipole orientations. 
        The H-dimer configuration, where two dipoles are aligned parallel to each other and positioned side-by-side, the interaction is positive, causing the bright state (Fig.~\ref{fig:system}) to lie energetically higher than the dark state.
        Conversely, in a J-dimer, where the dipoles are arranged in a head-to-tail configuration, the interaction is negative, leading to the dark state being energetically higher than the bright state.
        As the relative dipole orientation is varied between these limiting cases, there exists a critical ``magic angle” of $54.7^\circ$ between the dipole directions $\boldsymbol{\mu}_{1(2)}$ and the separation vector $\boldsymbol{r}_{1, 2}$ at which the dipole-dipole interaction vanishes. As a result, the bright and dark states become degenerate, so there is no phonon-assisted relaxation between them, and the optical decay rates depend only on the phonon-induced renormalisation of the energy levels.
        The relative positioning of the bright and dark states for the different dipole configurations affects the cascade through the single-excitation manifold when vibrations are introduced into the system.   

        The enhancement in the optical signals due to quantum correlations arising from entangled excitonic states and the ensuing quantum interference between emission pathways can be measured in a photon-counting experiment. 
        In this section, we model such a photon detection process by assuming point-like detectors in the far-field which pick up photons emitted along a specific direction $\boldsymbol{q}$.
        We calculate mode-resolved signatures to explore spatio-temporal effects in the intensity $I_{\boldsymbol{q}}(t)$ and photon correlations $g_{\boldsymbol{q}, \boldsymbol{q}^{\prime}}^{(2)}(\infty, \tau)$ in a Hanbury-Brows-Twiss (HBT) experiment.
        
        \subsection{Intensity}\label{subsec:int}
            The optical interaction Hamiltonian in Eq.~\eqref{eqn:H_opt} can be written as \cite{PhysRevA.107.023718}
            \begin{equation}\label{eqn:H_opt_RWA}
                \begin{aligned}
                    H_{\text{I,opt}} & = \sum_{\boldsymbol{q}, \lambda}  \mathcal{N}_{\boldsymbol{q}, \lambda}g_{\boldsymbol{q}} (\sigma_{\boldsymbol{q}, \lambda}^{+} a_{\boldsymbol{q}, \lambda} + \sigma_{\boldsymbol{q}, \lambda}^{-} a_{\boldsymbol{q}, \lambda}^{\dagger}),
                    \end{aligned}
            \end{equation}
            where 
            the coupling parameter $g_{\boldsymbol{q}} = \sqrt{\omega_{\boldsymbol{q}}/{2\epsilon_0 \hbar \mathcal{V}}}$.
            To describe the mode-selective coupling of the dipoles to the field, we define the effective raising (lowering) operator          
            \begin{equation}\label{eqn:raise_lower_op}
                \sigma_{\boldsymbol{q}, \lambda}^{\pm} = \frac{1}{\mathcal{N}_{\boldsymbol{q}, \lambda}} \left[\mu_{\boldsymbol{q}, \lambda}^{(1)}e^{\mp i\boldsymbol{q}\cdot \boldsymbol{r}/2} \sigma_{1}^{\pm} + \mu_{\boldsymbol{q}, \lambda}^{(2)}e^{\pm i\boldsymbol{q}\cdot \boldsymbol{r}/2} \sigma_{2}^{\pm}\right]
            \end{equation} 
            with normalisation $\mathcal{N}_{\boldsymbol{q}, \lambda} = \sqrt{|\mu_{\boldsymbol{q}, \lambda}^{(1)}|^{2} + |\mu_{\boldsymbol{q}, \lambda}^{(2)}|^{2}}$. 
            Here, $\mu_{\textbf{q}, \lambda}^{(m)} = \boldsymbol{\mu}_{m} \cdot \textbf{e}_{\textbf{q}, \lambda}$ (detailed in Appendix~\ref{sec:dip_proj}) is the projection of the $m\textsuperscript{th}$ dipole along the polarisation vector $\lambda$.
            We assume $\mu_{\textbf{q}, \lambda}^{(m)} \in \mathbb{R}$ and absorb the complex factor in Eq.~\eqref{eqn:H_opt} in the phase $e^{i \textbf{q}\cdot \textbf{r}}$.
            These operators describe transitions between different excitation manifolds mediated by emission or absorption of a photon along a specific direction $\textbf{q}$ and polarisation $\lambda$.
            
\begin{figure}[!ht]
\hspace{-0.5cm}
\begin{subfigure}[b]{0.43\textwidth}
        \centering
\begin{tikzpicture}
    \node[anchor=south west] at (-4.45, 1.5) {(a)};

    \node at (0.0, 1.8) (e1e2) {$\left| e_1, e_2 \right\rangle$};
    
    \fill[teal!20, draw=none, opacity=0.3] (-1.9, 0) ellipse (2.1 and 1.3);  
    \fill[teal!20, draw=none, opacity=0.3] (1.9, 0) ellipse (2.1 and 1.3);   
    
    \node at (-2.75, 0.35) (s1) {$| \psi_{\boldsymbol{q}_{1}^{\prime}}^{(g)} \rangle$};
    \node at (-2.0, 0.35) (s2) {$| \psi_{\boldsymbol{q}_{2}^{\prime}}^{(g)} \rangle$};
    \node at (-0.85, 0.35) (sn) {$| \psi_{\boldsymbol{q}_{n}^{\prime}}^{(g)} \rangle$};

    \node at (0.95, -0.3) (d1) {$| \psi_{\boldsymbol{q}_{1}}^{(e)} \rangle$};
    \node at (1.75, -0.3) (d2) {$| \psi_{\boldsymbol{q}_{2}}^{(e)} \rangle$};
    \node at (3.1, -0.3) (dn) {$| \psi_{\boldsymbol{q}_{n}}^{(e)} \rangle$};

    \node at (0, -1.8) (g1g2) {$\left| g_1, g_2 \right\rangle$};

    \draw[thick] (-1.0, 1.5) -- (1.0, 1.5);  
    \draw[thick] (-1.0, -1.5) -- (1.0, -1.5);  

    \draw[thick] (-3.2, 0) -- (-2.5, 0);  
    \draw[thick] (-2.4, 0) -- (-1.7, 0);  
    \draw[thick] (-1.2, 0) -- (-0.5, 0);  

    \draw[thick] (0.6, 0) -- (1.3, 0);    
    \draw[thick] (1.5, 0) -- (2.2, 0);    
    \draw[thick] (2.75, 0) -- (3.45, 0);  

    \node at (-1.4, 0.0) {$\cdots$};
    \node at (2.5, 0.0) {$\cdots$};
    \node at (3.7, 0.0) {$\cdots$};
    \node at (-3.5, 0.0) {$\cdots$};

    \draw[-Latex, thick, blue] (0.4, 1.3) -- (1.6, 0.2) node[anchor=west, midway, xshift=-2pt, yshift=2pt] {$\gamma_{\boldsymbol{q}_{2}}^{(e)}$}; 
    \draw[-Latex, thick, blue] (-1.9, -0.2) -- (-0.6, -1.3) node[anchor=west, midway, xshift=-2pt, yshift=2pt] {$\gamma_{\boldsymbol{q}_{2}^{\prime}}^{(g)}$};

    \draw[-Latex, thick, dashed, red] (0.0, 1.3) -- (0.9, 0.2);
    \draw[-Latex, thick, dashed, red] (1.1, 1.3) -- (3.0, 0.2);

    \draw[-Latex, thick, dashed, red] (-2.7, -0.2) -- (-1.1, -1.3);
    \draw[-Latex, thick, dashed, red] (-0.8, -0.2) -- (0.0, -1.3);

\end{tikzpicture}
\end{subfigure}

\begin{subfigure}[b]{0.22\textwidth}
        \centering
        \vspace{2mm}
\begin{tikzpicture}
    \node[anchor=south west] at (-2, 1.5) {(b)};

    \node at (0, 1.8) (e1e2) {$\left| e_1, e_2 \right\rangle$};
    \node at (1.1, 0.3) (psi2) {$| \psi_{\boldsymbol{q}}^{(e)} \rangle$};
    \node at (-1.0, 0.3) (psi2perp) {$| \psi_{\boldsymbol{q}}^{(e)\perp} \rangle$};
    \node at (0, -1.8) (g1g2) {$\left| g_1, g_2 \right\rangle$};

    \draw[thick] (-0.5, 1.5) -- (0.5, 1.5);       
    \draw[thick] (-1.5, 0) -- (-0.5, 0);          
    \draw[thick] (0.5, 0) -- (1.5, 0);     
    \draw[thick] (-0.5, -1.5) -- (0.5, -1.5);     
    
    \draw[-Latex, dashed] (-0.4, 1.2) -- (-1.0, 0.6);                         
    \draw[-Latex, thick] (0.4, 1.3) -- node[right] {$\gamma_{\boldsymbol{q}}^{(e)}$} (0.8, 0.6); 
    
\end{tikzpicture}
    \end{subfigure}
    \hspace{2mm}
        \begin{subfigure}[b]{0.22\textwidth}
        \centering
        \vspace{5mm}

        \begin{tikzpicture}
    \node[anchor=south west] at (-2, 1.5) {(c)};

    \node at (0, 1.8) (e1e2) {$\left| e_1, e_2 \right\rangle$};
    \node at (1.1, -0.3) (psi1) {$| \psi_{\boldsymbol{q}^{\prime}}^{(g)} \rangle$};
    \node at (-1.1, -0.3) (psi1perp) {$| \psi_{\boldsymbol{q}^{\prime}}^{(g)\perp} \rangle$};
    \node at (0, -1.8) (g1g2) {$\left| g_1, g_2 \right\rangle$};

    \draw[thick] (-0.5, 1.5) -- (0.5, 1.5);       
    \draw[thick] (-1.5, 0) -- (-0.5, 0);          
    \draw[thick] (0.5, 0) -- (1.5, 0);     
    \draw[thick] (-0.5, -1.5) -- (0.5, -1.5);     

    \draw[-Latex, thick] (1.0, -0.7) -- node[right] {$\gamma_{\boldsymbol{q}^{\prime}}^{(g)}$} (0.3, -1.3); 
    \draw[-Latex, dashed] (-1.0, -0.7) -- (-0.4, -1.3);                       
\end{tikzpicture}
    \end{subfigure}

\begin{subfigure}[b]{0.45\textwidth}
    \centering
    \vspace{3mm}
    \begin{tikzpicture}
        \node[anchor=south west, inner sep=0] (image) at (-2, -1) {\includegraphics[width= \textwidth]{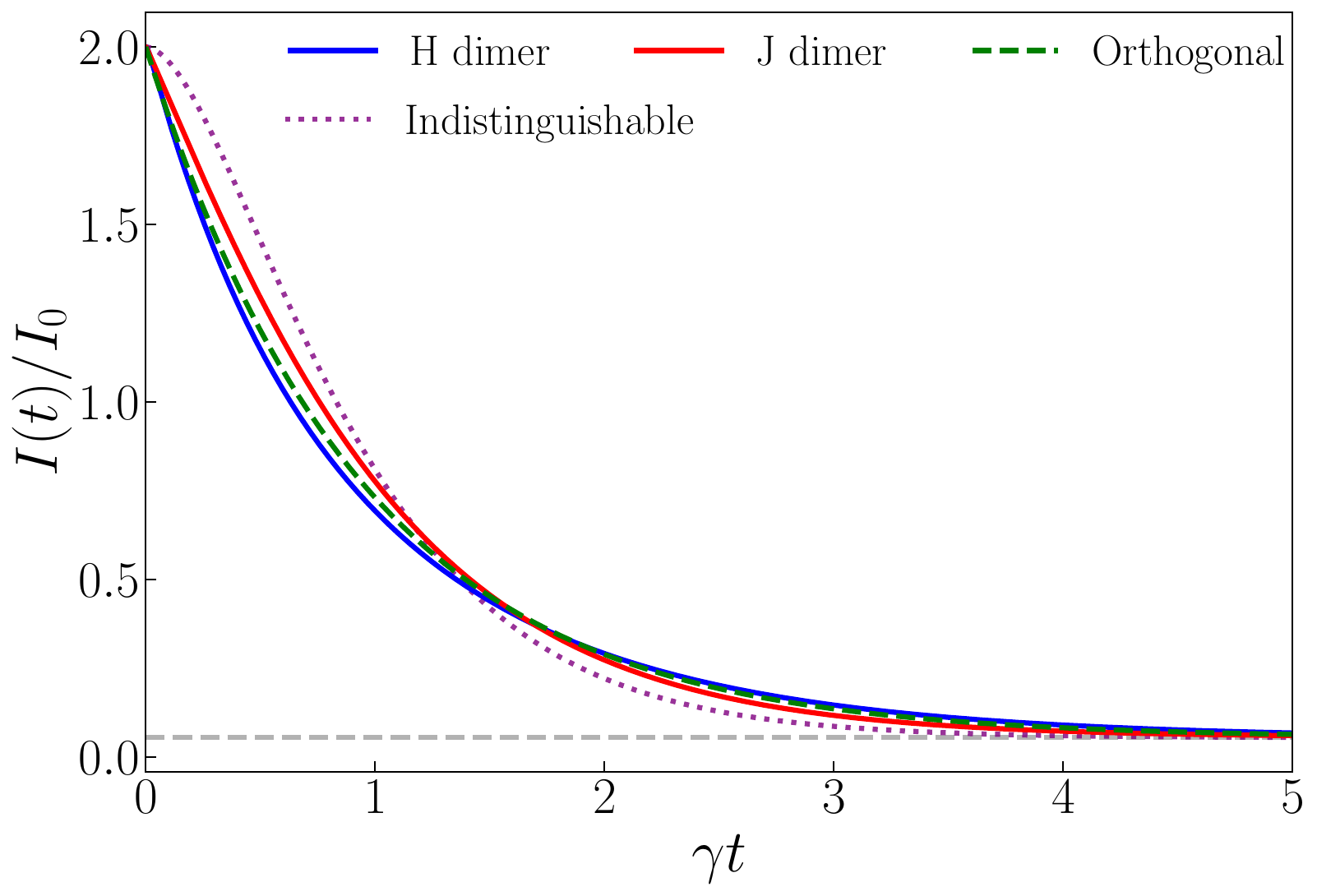}};
        \node[anchor=north west] at ([xshift=-6pt, yshift=-1pt] image.north west) {(d)};
        \tikzset{
            state/.style={inner sep=1pt, anchor=center}, 
            line/.style={thick}, 
            arrow/.style={-Latex, thick}, 
            dashedarrow/.style={Latex-, dashed, thick}, 
            ovalstyle/.style={draw, thick} 
        }
        \begin{scope}[shift={(0.7, 1)}] 
            \node[anchor=south west, scale=0.7] at (0, -0.5) {
                \begin{tikzpicture}[font=\normalsize] 
                    \node[state] at (-1.0, 1.8) (e1e2) {$\left| e_1, e_2 \right\rangle$};
                    \node[state] at (-2.5, 0.75) (psiA) {$| \psi_{D} \rangle$};
                    \node[state] at (0.1, 0.25) (psiS) {$| \psi_{B} \rangle$};
                    \node[state] at (-1.0, -1.8) (g1g2) {$\left| g_1, g_2 \right\rangle$};

                    \draw[line] (-1.5, 1.5) -- (-0.5, 1.5);
                    \draw[line] (-1.5, -1.5) -- (-0.5, -1.5);
                    \draw[line] (-2.5, 0.5) -- (-1.5, 0.5);
                    \draw[line] (-1.0, 0) -- (0.0, 0);

                    \draw[arrow] (-0.8, 1.2) -- node[right] {$\gamma^{1}_{B}$} (-0.5, 0.4);
                    \draw[arrow] (-0.5, -0.4) -- node[right] {$\gamma^{2}_{B}$} (-0.8, -1.2);
                    \draw[dashedarrow] (-1.8, 0.35) -- (-0.7, 0.1);
                \end{tikzpicture}
            };
            \draw[ovalstyle, red] (1.23, 1.0) ellipse (1.25cm and 1.5cm); 
        \end{scope}

        \begin{scope}[shift={(3.4, 0.6)}] 
            \node[anchor=south west, scale=0.7] at (0, 0) {
                \begin{tikzpicture}[font=\normalsize] 
                    \node[state] at (-1.0, 1.8) (e1e2) {$\left| e_1, e_2 \right\rangle$};
                    \node[state] at (-2.5, -0.25) (psiA) {$| \psi_{D} \rangle$};
                    \node[state] at (0.1, 0.25) (psiS) {$| \psi_{B} \rangle$};
                    \node[state] at (-1.0, -1.8) (g1g2) {$\left| g_1, g_2 \right\rangle$};

                    \draw[line] (-1.5, 1.5) -- (-0.5, 1.5);
                    \draw[line] (-1.5, -1.5) -- (-0.5, -1.5);
                    \draw[line] (-2.5, -0.5) -- (-1.5, -0.5);
                    \draw[line] (-1.0, 0) -- (0.0, 0);

                    \draw[arrow] (-0.8, 1.2) -- node[right] {$\gamma^{1}_{B}$} (-0.5, 0.4);
                    \draw[arrow] (-0.5, -0.4) -- node[right] {$\gamma^{2}_{B}$} (-0.8, -1.2);
                    \draw[dashedarrow] (-1.8, -0.35) -- (-0.7, -0.1);
                \end{tikzpicture}
            };
            \draw[ovalstyle, blue] (1.25, 1.5) ellipse (1.22cm and 1.5cm); 
        \end{scope}

        \draw[arrow, red] (0.98, 2.95) .. controls (0.5, 3.0) .. (-0.5, 2.8); 
        \draw[arrow, blue] (3.82, 1.0) .. controls (3.7, 0.7) .. (3.7, 0.11); 
    \end{tikzpicture}
\end{subfigure}
    \caption{\justifying Illustration of decay pathways under mode-selective photon detection in optical modes $\boldsymbol{q}$ and $\boldsymbol{q}^{\prime}$. 
    Blue arrows indicate transitions involving specific intermediate states $|\psi_{\boldsymbol{q}}^{(e)}\rangle$ and $|\psi_{\boldsymbol{q}^{\prime}}^{(g)}\rangle$, onto which the system is projected after photon emission. 
    Unrealized decay channels are shown as red dashed arrows.
    These sets of intermediate states depend on the relative orientation of dimer dipoles and are equivalent only if the dipoles are identical.
    Panels (b) and (c) show energy level diagrams for transitions from the doubly excited state to the single-excitation manifold, and from there to the ground state, under detection of specific $\boldsymbol{q}$- and $\boldsymbol{q}^{\prime}$-modes. 
    In contrast to (a), these are shown from the reduced system's perspective in the space of electronic states. 
    Panel (d) shows emission intensity over time for different dipole orientations. 
    Insets highlight level structures for H- and J-dimer configurations, marked by blue and red circles, respectively.}
    \label{fig:orient_int}

\end{figure}

            We illustrate the level scheme for such directional emission in Fig.~\ref{fig:orient_int}(a), by introducing $\mu_{\textbf{q}, \lambda}^{(m)}$-dependent intermediate states whose form depends on both the emission direction $\textbf{q}$ and polarisation $\lambda$. 
            The raising and lowering operators in Eq.~\eqref{eqn:raise_lower_op} can be rewritten in terms of these states as
            \begin{equation}\label{eqn:ralo_op}
                \begin{aligned}
                    \sigma_{\textbf{q}, \lambda}^{+} &= |\psi^{(g)}_{\textbf{q}, \lambda}\rangle\bra{g_{1}g_{2}} + \ket{e_{1}e_{2}}\langle \psi^{(e)}_{\textbf{q}, \lambda} |, \\
                    \sigma_{\textbf{q}, \lambda}^{-} &= \ket{g_{1}g_{2}}\langle \psi^{(g)}_{\textbf{q}, \lambda}| + |\psi^{(e)}_{\textbf{q}, \lambda} \rangle \bra{e_{1}e_{2}}.
                \end{aligned}
            \end{equation}
            Here, the two families of intermediate states $\{|\psi_{\textbf{q}n}^{(e)}\rangle\}$ and $\{|\psi_{\textbf{q}^{\prime}n}^{(g)}\rangle\}$ define all possible decay channels, assuming the two photons are emitted along different directions $\textbf{q}$ and $\textbf{q}^{\prime}$ respectively \cite{PhysRevA.107.023718}.
            Each such state is a normalised superposition
            \begin{equation}
                |\psi_{\textbf{q}, \lambda}^{(g, e)} \rangle = \left(\mu_{\textbf{q}, \lambda}^{(1, 2)} \ket{e_{1} g_{2}} + \mu_{\textbf{q}, \lambda}^{(2, 1)}  \ket{g_{1} e_{2}}\right) / \mathcal{N}_{\textbf{q}, \lambda},
            \end{equation}
            where we have dropped the global phase factors $\exp\left[\pm i\textbf{q}\cdot \textbf{r} / 2 \right]$ under the sub-wavelength approximation $(\textbf{q} \cdot \textbf{r} \to 0)$ \cite{Rouse_2019}. 
            Consider a point-like detector placed in the far field of the dimer. Then, a detector click implies a position measurement of a photon. Far-field optics predicts that only a selected set of photon modes are picked up by the detectors, namely those with emission directions strongly focussed around the emitter-to-detector direction $\textbf{d}$. These selected photon modes couple to the the emitters with a common phase as described by the operator $\sigma_{\textbf{q}, \lambda}^{\pm}$, where $\textbf{q}$ is a reference wave vector parallel to $\textbf{d}$. 
            A detection event then corresponds to a projection onto a state which is an eigenstate of the measurement operator $\sigma_{\boldsymbol{q}, \lambda}^{\pm}$, associated with that emission direction.

            For a given dimer configuration, the direction of photon detection --- along wave vector $\boldsymbol{q}$ and $\boldsymbol{q}^{\prime}$, as shown in Fig.~\ref{fig:orient_int}(b) and (c) --- thus selectively defines the decay channels via the intermediate states $| \psi_{\boldsymbol{q}, \lambda}^{(e)} \rangle$ and $| \psi_{\boldsymbol{q}^{\prime}, \lambda}^{(g)} \rangle$.
            The state $| \psi_{\boldsymbol{q}}^{(e)} \rangle$ channels the excitations from the doubly excited state whereas $| \psi_{\boldsymbol{q}^{\prime}}^{(g)} \rangle$ facilitates transitions from the single excitation manifold into the ground state.
            For simplicity, we omit the polarization index here and in Fig.~\ref{fig:orient_int}, as it plays no essential role in the directional selectivity under consideration.
                      
            In case of a parallel dimer, where the dipole projections are equal, $\mu_{\boldsymbol{q}, \lambda}^{(1)} = \mu_{\boldsymbol{q}, \lambda}^{(2)}$, the intermediate states for photon emission in any direction and polarisation simplify to the bright state, i.e., $| \psi_{\boldsymbol{q}, \lambda}^{(g)} \rangle = | \psi_{\boldsymbol{q}, \lambda}^{(e)} \rangle = | \psi_{B} \rangle$. 
            Conversely, for non-parallel dipoles, where $\mu_{\boldsymbol{q}, \lambda}^{(1)} \neq \mu_{\boldsymbol{q}, \lambda}^{(2)}$, the projections of the dipole vectors onto the polarization directions become crucial as detailed in Sec.~\ref{subsec:ph_coin}, where the cooperative effects in orthogonal dipoles ($\mu_{\boldsymbol{q}, \lambda}^{(1)} \neq \mu_{\boldsymbol{q}, \lambda}^{(2)}$) depend on the light field mode $\boldsymbol{q}$ being collected.
            However, the overall collective emission rate is determined by the combined contributions from individual decay processes across all collected modes \cite{PhysRevA.107.023718}. 
            In the case of an orthogonal dimer, this reduces to decay into the site basis states.

            In Appendix~\ref{sec:dir_int}, we derive in detail the direction-dependent intensity characterised by these decay channels. Assuming weak dipole-dipole coupling, the expression simplifies to
            \begin{equation}\label{eqn:dir_int}
                I_{\boldsymbol{q}}(t) = \sum_{\lambda = 1}^{2} 2\pi g_{\boldsymbol{q}}^{2} \mathcal{N}_{\boldsymbol{q}, \lambda}^{2} \delta(\omega_{\boldsymbol{q}} - \omega) \langle \sigma_{\boldsymbol{q}, \lambda}^{+}(t) \sigma_{\boldsymbol{q}, \lambda}^{-}(t) \rangle,
            \end{equation}
            which links the observed radiation pattern to the quantum state of the system. 
            Then, defining the occupation of the excitonic states $\ket{e_{1}e_{2}}$, $\ket{e_{1}g_{2}}$ and $\ket{g_{1}e_{2}}$ as $n_{e_{1} e_{2}}$, $n_{e_{1} g_{2}}$ and $n_{g_{1} e_{2}}$, respectively, the total intensity can be derived by integrating over all accessible light-modes and can be expressed as,
            \begin{multline}\label{eqn:tot_int}
                    I (t) = \gamma_{1} n_{e_{1} g_{2}} + \gamma_{2} n_{e_{2} g_{1}} + (\gamma_{1} + \gamma_{2}) n_{e_{1} e_{2}}\\
                    + \sqrt{\gamma_{1}\gamma_{2}} \mathcal{F} (\text{Tr}\left[\ket{g_{1}e_{2}}\bra{e_{1}g_{2}}\rho\right] + \text{Tr}\left[\ket{e_{1}g_{2}}\bra{g_{1}e_{2}}\rho \right]),
            \end{multline}
            where $\mathcal{F} = \boldsymbol{\mu}_{1} \cdot \boldsymbol{\mu}_{2}$ is a cross-function \cite{ficek2005quantum, Rouse_2019} and the decay rate from each of the energy eigenstates is, $\gamma_{m} = \omega^{3} \mu_{m}^{2} / (3 \pi \epsilon_{0} \hbar c^{3})$.

            We plot the time-dependent intensity for three different geometric configurations of dimers in Fig.~\ref{fig:orient_int}(d).
            The purple dotted line, shows the idealised case when two uncoupled quantum emitters emit collectively without vibrational effects, resulting in a distinct non-exponential dynamics in the intensity profile. 
            H- and J-dimers, on the other hand, show different intensity profiles owing to phonon-assisted excitation and relaxation between the bright and the dark states, as shown in the insets circled in blue and red, respectively, in Fig.~\ref{fig:orient_int}(d).
            The bright state population of the J-dimer is higher than that of the H-dimer at early times, having its dark state energetically higher and less accessible.
            Overall, the J-dimer features an enhanced decay rate, which may be reduced from the ideal superradiant case of two idealised identical emitters if vibrational excitation becomes relevant, or in the presence of strong vibrational coupling also at zero temperature \cite{PhysRevResearch.6.033231}. 
            By contrast, the intensity of the H-dimer rapidly decays, owing to vibrationally assisted transition of excitations from the bright to the dark state. 
            Finally, for orthogonal dipoles the contribution from the coherence terms in Eq.~\eqref{eqn:tot_int} vanishes owing to the cross-function $\mathcal{F}$ \cite{ficek2005quantum, Rouse_2019} going to zero. 
            Thus, upon integrating over all light modes, excitations follow two distinct channels (one for each site basis state), at the single emitter decay rate $\gamma$.
        \subsection{Photon Coincidences} \label{subsec:ph_coin}
            Measuring mode-resolved quantities like the correlation between photons sampled along particular directions give us additional insights into other cooperative effects besides superradiance in a two-emitter system.
            The photon coincidence for two subsequent detections (along $\boldsymbol{q}$ and $\boldsymbol{q}^{\prime}$) can be written in terms of system operators as follows \cite{PhysRevA.107.023718},
            \begin{equation}\label{eqn:g2}
            \begin{split}
            G^{(2)}_{\boldsymbol{q}, \boldsymbol{q}^{\prime}}(t, \tau) 
            &= \sum_{\lambda, \lambda^{\prime} = 1}^{2}
                \langle I_{\boldsymbol{q}, \lambda}(t) I_{\boldsymbol{q}^{\prime}, \lambda^{\prime}}(t + \tau) \rangle \\
            &\propto \sum_{\lambda, \lambda^{\prime} = 1}^{2}
                \langle \sigma_{\boldsymbol{q}, \lambda}^{+}(0) \sigma_{\boldsymbol{q}^{\prime}, \lambda^{\prime}}^{+}(\tau) 
                \sigma_{\boldsymbol{q}^{\prime}, \lambda^{\prime}}^{-}(\tau) \sigma_{\boldsymbol{q}, \lambda}^{-}(0) \rangle, \\
            g^{(2)}_{\boldsymbol{q}, \boldsymbol{q}^{\prime}}(t, \tau)
            &= \frac{
                G^{(2)}_{\boldsymbol{q}, \boldsymbol{q}^{\prime}}(t, \tau)
            }{
                \sum_{\lambda, \lambda^{\prime} = 1}^{2}
                \langle I_{\boldsymbol{q}, \lambda}(t) \rangle 
                \langle I_{\boldsymbol{q}^{\prime}, \lambda^{\prime}}(t + \tau) \rangle
            }.
            \end{split}
            \end{equation}
            where $G^{(2)}_{\boldsymbol{q}, \boldsymbol{q}^{\prime}}(t, \tau)$ is the unnormalised and $g^{(2)}_{\boldsymbol{q}, \boldsymbol{q}^{\prime}}(t, \tau)$ is the normalised two-photon intensity correlation \cite{ficek2005quantum}. 
            Upon expanding the mode-resolved ladder operators in Eq.~\eqref{eqn:ralo_op}, correlations of the form $\langle \sigma_{m}^{+}(t) \sigma_{n}^{+}(t^{\prime}) \sigma_{n}^{-}(t^{\prime}) \sigma_{m}^{-} (t) \rangle$ in Eq.~\eqref{eqn:g2} describe the probabilities of detecting photons emitted from either the same $(m = n)$ or different $(m \neq n)$ emitters. 
            Specifically, for $m \neq n$, these correlations are proportional to the likelihood of detecting a photon at time $t^{\prime}$ from the $n\textsuperscript{th}$ emitter, given that a photon from the $m\textsuperscript{th}$ emitter was detected at time $t$.
            Furthermore, these photon coincidences also involve dipole correlations of the form $\langle \sigma_{m}^{+}(t) \sigma_{n}^{+}(t^{\prime}) \sigma_{m}^{-}(t^{\prime}) \sigma_{n}^{-}(t) \rangle$, which arise due to correlated photons emitted from coherent superpositions of emitter eigenstates.          
            
            We first focus on the unnormalised $G^{(2)}_{\boldsymbol{q}, \boldsymbol{q}^{\prime}}(t, \tau)$ at steady-state by taking the limit $t \to \infty$. 
            We examine how its value at zero-time delay $(\tau = 0)$ depends on the overlap between the intermediate decay states $|\psi^{i = (g, e)}_{\boldsymbol{q}, \lambda} \rangle$ [as shown in Fig.~\ref{fig:orient_int}(a)], such that
            \begin{equation}
                \begin{aligned}\label{eqn:G2}
                 G^{(2)}_{\boldsymbol{q}, \boldsymbol{q}^{\prime}}(\infty, 0) & =\langle \sigma^{+}_{\boldsymbol{q}, \lambda}(0)\sigma^{+}_{\boldsymbol{q}^{\prime}, \lambda^{\prime}}(0) \sigma^{-}_{\boldsymbol{q}^{\prime}, \lambda^{\prime}}(0) \sigma^{-}_{\boldsymbol{q}, \lambda}(0)\rangle \\
                 & = \langle|\langle \psi^{(e)}_{\boldsymbol{q}, \lambda} |\psi^{(g)}_{\boldsymbol{q}^{\prime}, \lambda^{\prime}} \rangle|^{2} \ket{e_{1}e_{2}} \bra{e_{1}e_{2}}\rangle.
                \end{aligned}
            \end{equation}
            Next, we assume that the two subsequent detections, $\boldsymbol{q}$ and $\boldsymbol{q}^{\prime}$, shown in Fig.~\ref{fig:system}(b), are associated with spherical angles $(\theta, \theta^{\prime})$ and azimuthal angles $(\phi, \phi^{\prime})$.
            Substituting the dipole projections $\mu_{\boldsymbol{q}, \lambda}^{(m)}$ (calculated in Appendix~\ref{sec:dip_proj}) in Eq.~\eqref{eqn:G2}, we can write down the steady-state $G^{(2)}_{\boldsymbol{q}, \boldsymbol{q}^{\prime}}(\infty, 0)$ at zero-delay for H- and J-dimers as,
            \begin{equation}\label{eqn:G2_HJ}
                \begin{aligned}
                    G^{(2)}_{\text{H-dimer}} &= 4 n_{e_{1}e_{2}}(\cos^{2} \theta \cos^{2} \phi + \sin^{2} \phi) \\
                    & (\cos^{2} \theta^{\prime} \cos^{2} \phi^{\prime} + \sin^{2} \phi^{\prime}) 
                    \\
                    G^{(2)}_{\text{J-dimer}} & = 4 n_{e_{1}e_{2}}\sin ^2 \theta \sin ^2 \theta^{\prime} 
                    ,
                \end{aligned}
            \end{equation}
            where the angular factors for the different dimer configurations give rise to regions of constructive and destructive interference in the radiation pattern. 
            
            We now calculate the normalised photon coincidences for H- and J-dimers using mode-selective detectors which capture photons emitted along the same direction (i.e., $\boldsymbol{q} = \boldsymbol{q}^{\prime}$).
            For closely spaced molecular dimers selectively exciting individual sites is generally not feasible due to strong near-field interactions and significant overlap in their optical responses \cite{doi:10.1021/jp907644h, 10.1063/5.0064133}. 
            Thus, in this study, we incoherently pump the bright state (with most optical activity) with rate $\gamma_{p}$ matched to that of the optical decay, to have homogeneity in our study \cite{PhysRevResearch.6.023207}.
            In Sec.~\ref{subsec:int}, we have observed that optical decay is direction dependent, which conversely entails that we can populate the symmetric superposition of site-basis states for different dimers by pumping along different directions
            \footnote{For example, for parallel dipoles, pumping along any direction drives the bright state, which is an eigenstate of the system. 
            For orthogonal emitters, we have found in Sec.~\ref{subsec:int}, integrating over all possible modes results in two distinct channels of site basis states.
            However, Eq.~\eqref{eqn:ralo_op} implies that measuring along $\boldsymbol{q}=(1, -1, 0)$ helps us access the symmetric state $(|e_1 g_2\rangle + |g_1 e_2\rangle) / \sqrt{2}$.}.

            \begin{figure}[htbp]
            \begin{subfigure}{0.48\textwidth}
                \centering
                \begin{overpic}[width=\textwidth]{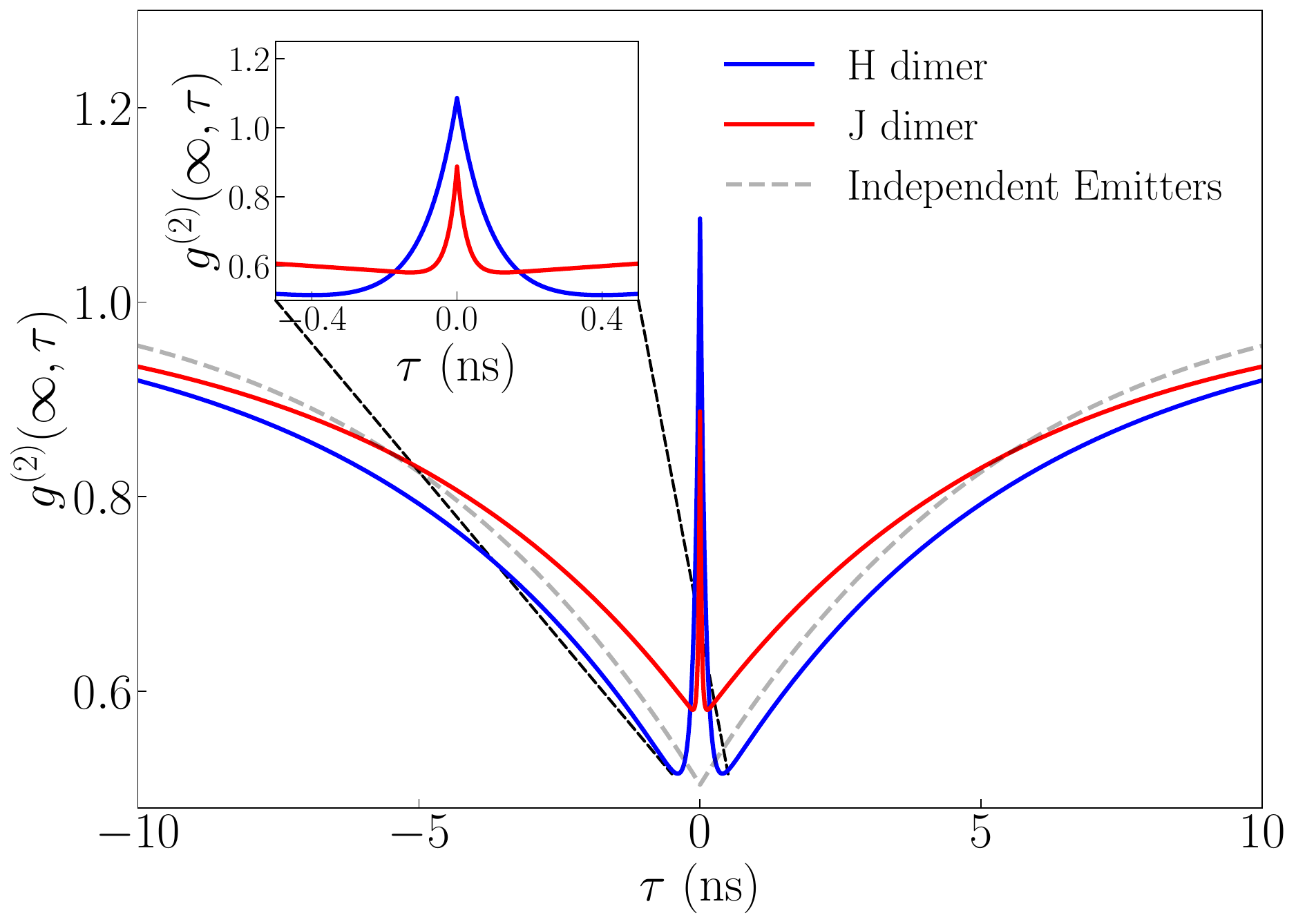} 
            \put(0,65){
            \begin{tikzpicture}
            \node[anchor=north west] at (0, 100) {(a)};
            \end{tikzpicture}}
            \end{overpic} 
            \end{subfigure}
            \begin{subfigure}{0.22\textwidth}
            \centering
            \begin{overpic}[width=\textwidth]{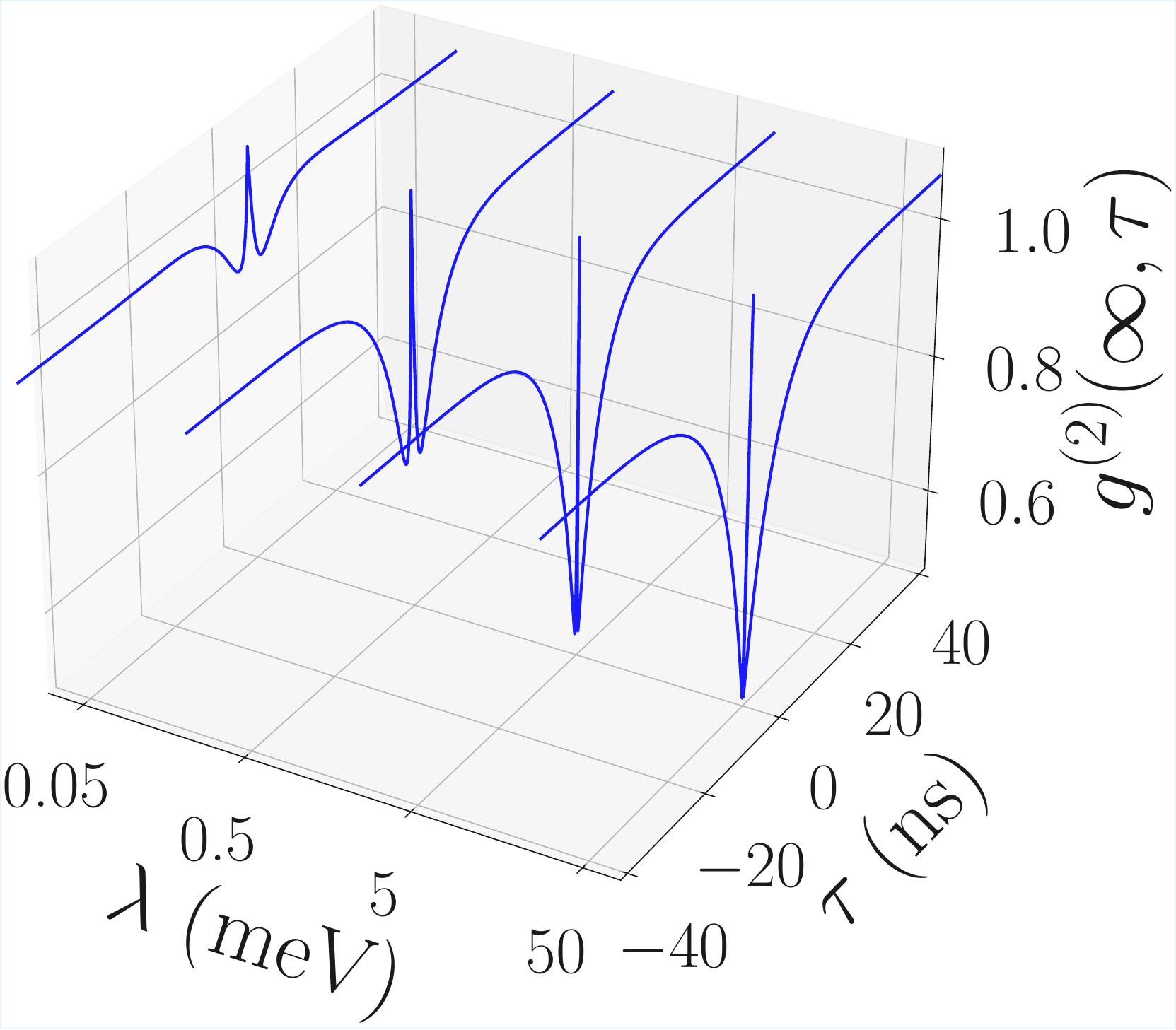} 
            \put(-1.5,75){
            \begin{tikzpicture}
            \node[anchor=north west] at (0, 100) {(b)};
            \end{tikzpicture}}
            \end{overpic}
            \end{subfigure}
            \hspace{3mm}
            \begin{subfigure}{0.22\textwidth}
                \centering
                \begin{overpic}[width=\textwidth]{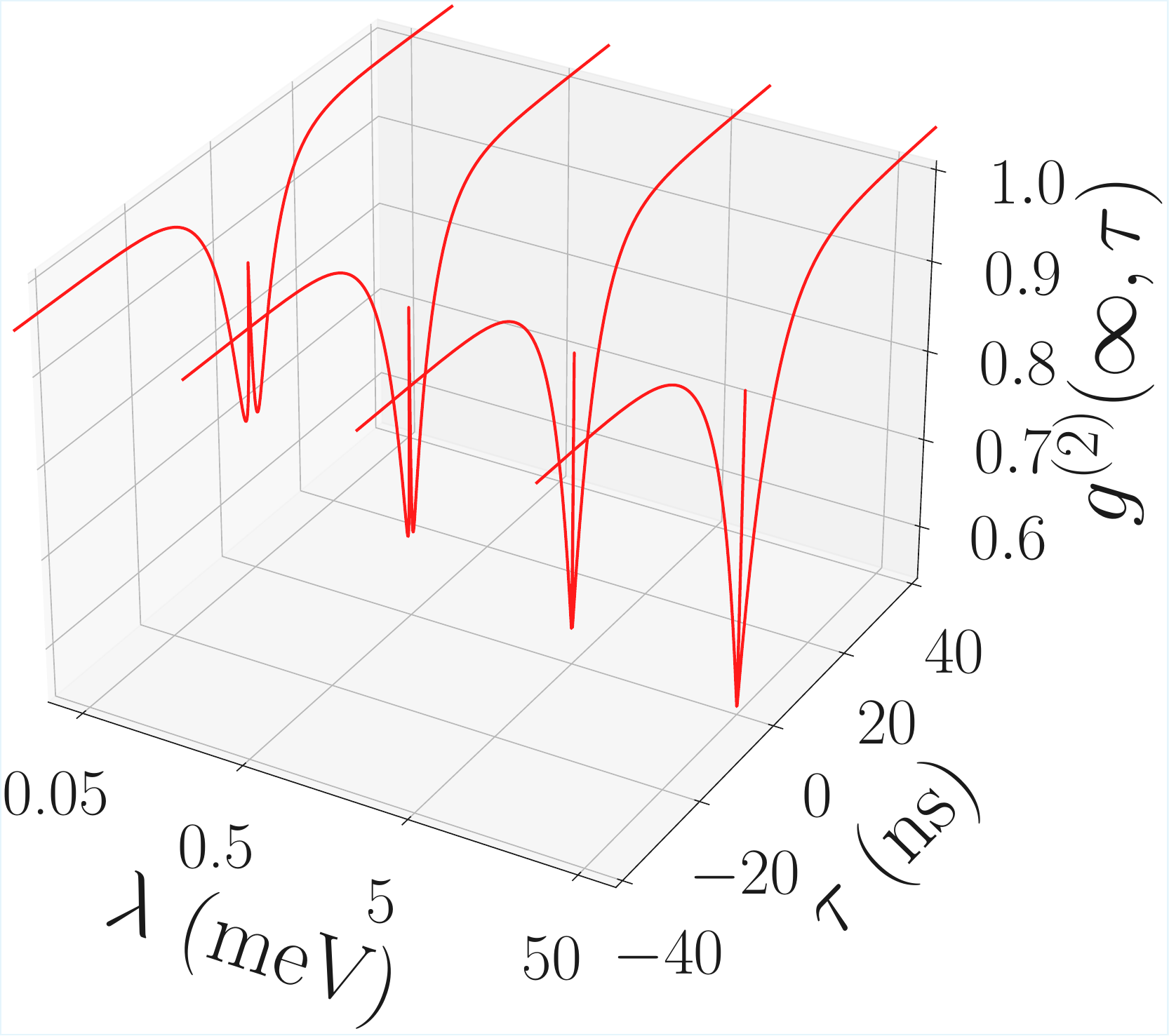} 
            \put(-1.5,75){
            \begin{tikzpicture}
            \node[anchor=north west] at (0, 100) {(c)};
            \end{tikzpicture}}
            \end{overpic}
            \end{subfigure}
            \caption{\justifying (a) Two-photon coincidences for different dimer configurations with pumping rate $\gamma_{p} = \gamma$. (b and c) Photon coincidence for H- and J-dimers plotted for different reorganisation energies of the phonon bath.}
                \label{fig:orient_g2}
            \end{figure}  
            
            Fig.~\ref{fig:orient_g2} shows the photon coincidence $g_{\boldsymbol{q}, \boldsymbol{q}^{\prime}}^{(2)}(\infty, \tau)$ for H- and J-dimers as a function of time delay $\tau$.
            We see that for such parallel dipoles, the $\boldsymbol{q}$-resolved photon correlations show an anti-dip at a zero time-delay.
            For parallel dipoles, $g^{(2)}(\infty, \tau)$ does not depend on the photon mode being detected because even though the photons emitted along other directions are less intense, they are equally correlated.
            Note that, henceforth, unless necessary, we have dropped the mode-index ($\boldsymbol{q}, \boldsymbol{q}^{\prime}$), since we assume both photons are detected along the same direction $\boldsymbol{q} = \boldsymbol{q}^{\prime}$. 
            This holds except for the case when detectors are placed parallel to the dipoles, along which the dimer does not emit. 
            We also find that the zero-delay $g^{(2)}(\infty, 0)$ is different for different dimer configurations, owing to the different populations of the system eigenstates at steady state. 
            
            When both the photons are detected perpendicular to the dipole orientations, we can arrive at an analytical expression for the normalised photon coincidence from  Eq.~\eqref{eqn:g2} and~\eqref{eqn:G2}. 
            For a parallel dimer $g^{(2)} (\infty, 0) =  n_{ee}/(n_{\text{ee}} + n^{\boldsymbol{q}}_{S})^{2}$, where $n^{\boldsymbol{q}}_{S}$ is the population of the $\boldsymbol{q}$-dependent bright state.  
            We have discussed the population dynamics of the bright and dark state of the two dimers in Sec.~\ref{subsec:int}, and observed that the excitations get trapped in the dark state of the H-dimer.
            Thus, at steady state, a larger number of excitations go into the dark state, causing the population of the bright state to plummet, leading to $g^{(2)}(\infty, 0) > 1$. 
            On the other hand, the relative inaccessibility of the dark state for the J-dimer leads to a higher steady-state bright state population, and hence the normalised $g^{(2)}(\infty, 0)$ has a lower anti-dip.
            Under the weak pumping ($\gamma_p$) and decay ($\gamma$) conditions considered here --- both much smaller than the phonon dephasing rate --- the second photon emission predominantly originates from the single-excitation manifold.
            Thus, at short time delays, the value of the photon coincidence depends on the relative population of the bright and dark states which then recovers back to a steady-state value at longer times, giving $g^{(2)}(\infty, \tau) = 1$ for $\tau \to \infty$. 
            We also observe a broader anti-dip for the H-dimer at short time delays, owing to a lower residual population in the bright state at short time delays. 
            
            Fig. \ref{fig:orient_g2}(b) and (c) further show the effect of strong phonon coupling on inter-emitter coherence. Increasing the phonon reorganization energy leads to faster coherence decay, resulting in a significantly narrower envelope of \(g^{(2)}(\infty, \tau)\). This general trend aligns with prior studies of molecular aggregates under strong system-bath coupling, where coherence lifetimes are known to be rapidly suppressed in the presence of large reorganization energies and overdamped bath dynamics \cite{tempelaar2014mapping}.
            
            Interestingly, the height of the anti-dip at zero-time delay stays constant. 
            This is because the pumping rate is the same as the optical decay rate, which causes the formation of a stationary state under an ``effective" infinite temperature photon bath. 
            The zero-delay $g^{(2)}(\infty, 0)$, which is fully determined by the stationary state and the measurement operators and does not depend on the system dynamic, thus remains constant. 
            
            \tdplotsetmaincoords{70}{110} 
            \begin{figure}[h]  
            
              \begin{subfigure}[h]{\columnwidth} 
                \begin{overpic}[width=0.48\textwidth]{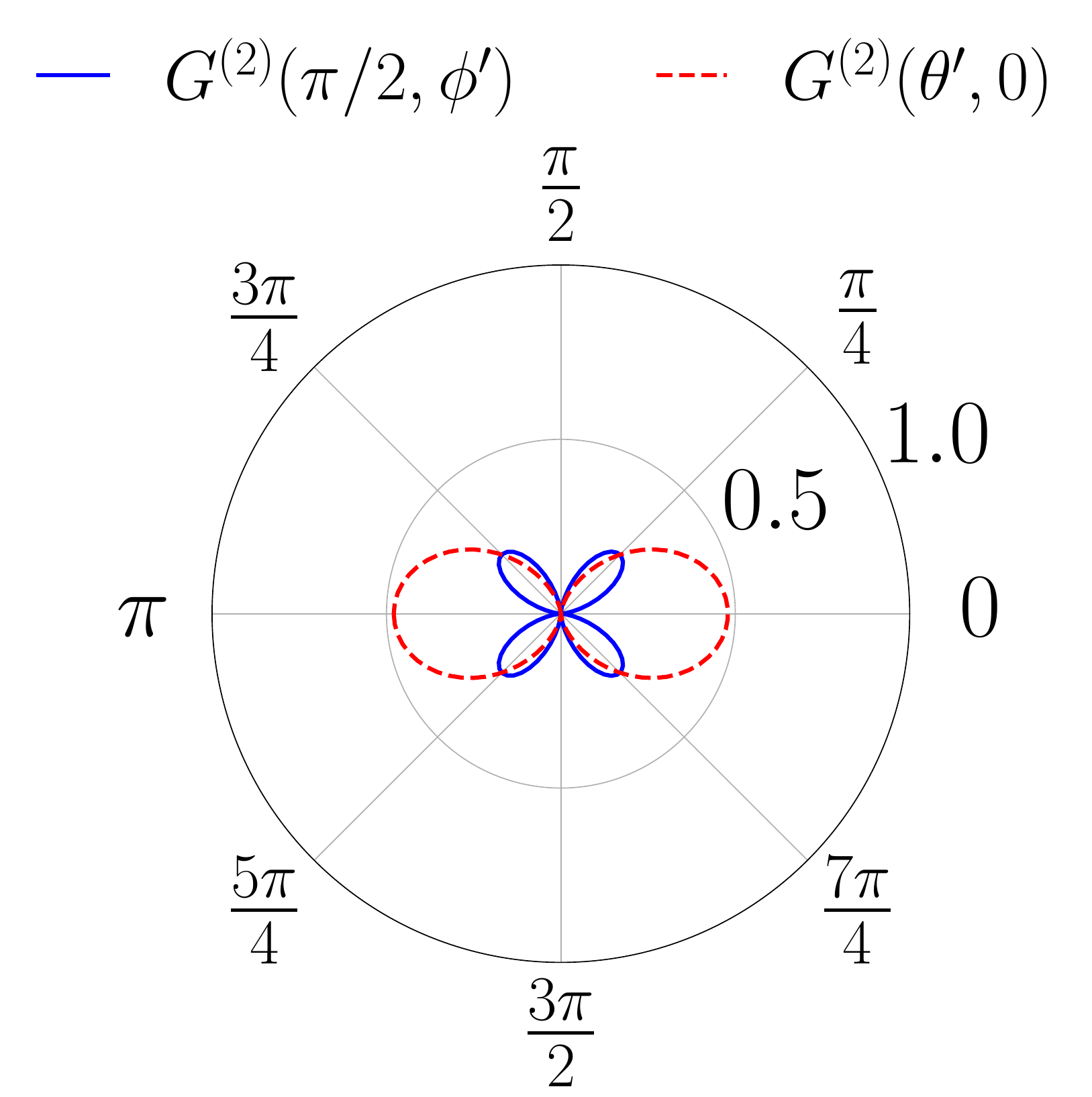} 
                            \put(-1,76){
                            \begin{tikzpicture}
                            \node[anchor=north west] at (0, 100) {(a)};
                            \end{tikzpicture}}
                            \end{overpic} 
                \hfill 
                \begin{overpic}[width=0.48\textwidth]{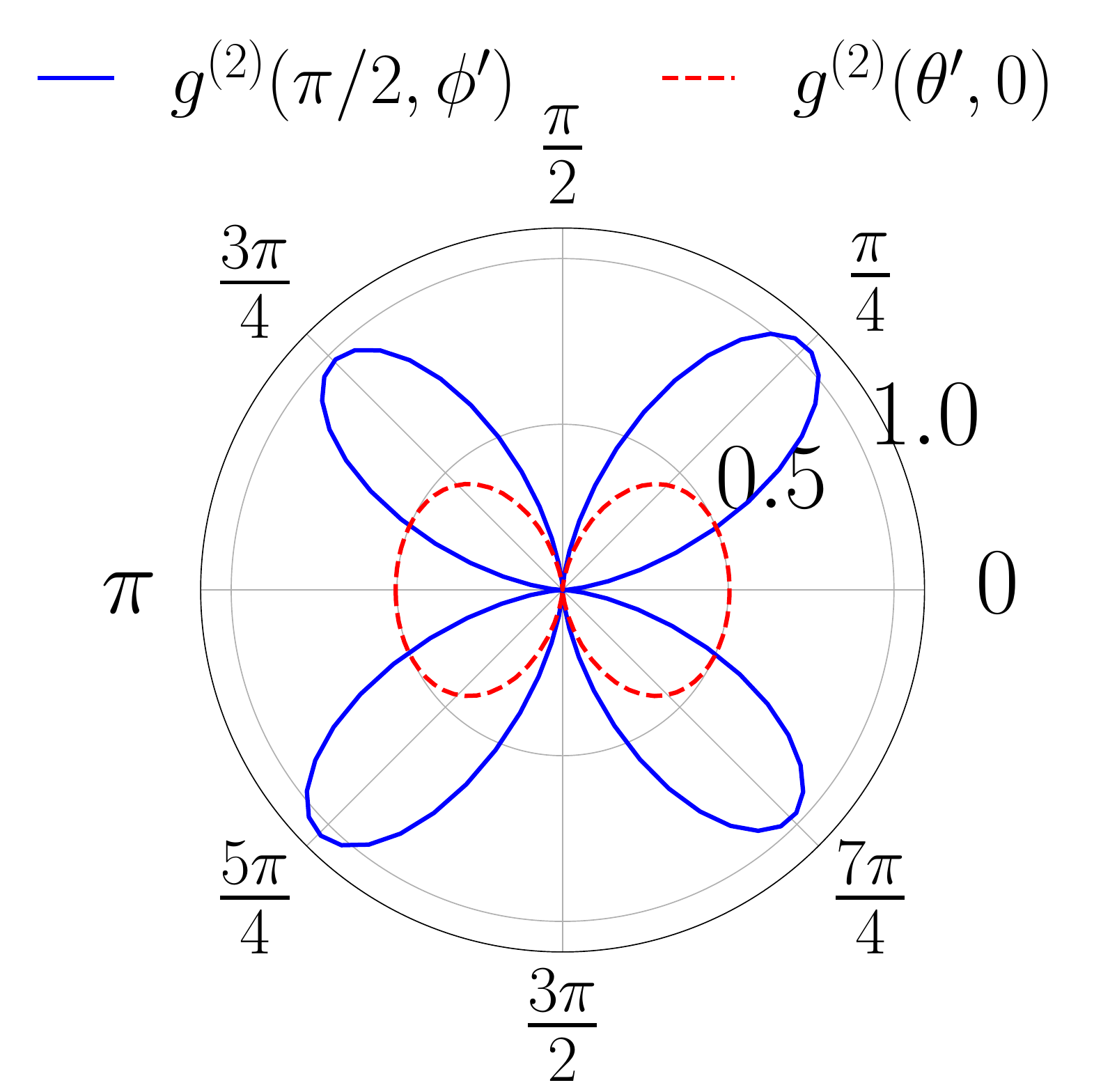} 
                \put(0,76){
                \begin{tikzpicture}
                \node[anchor=north west] at (0, 100) {(b)};
                \end{tikzpicture}}
                \end{overpic} 
                \vspace{0.01\textwidth} 
                \label{fig:subfig3} 
              \end{subfigure} 
              \hfill

              \begin{subfigure}[h]{0.9\columnwidth} 
                \begin{overpic}[width=\textwidth]{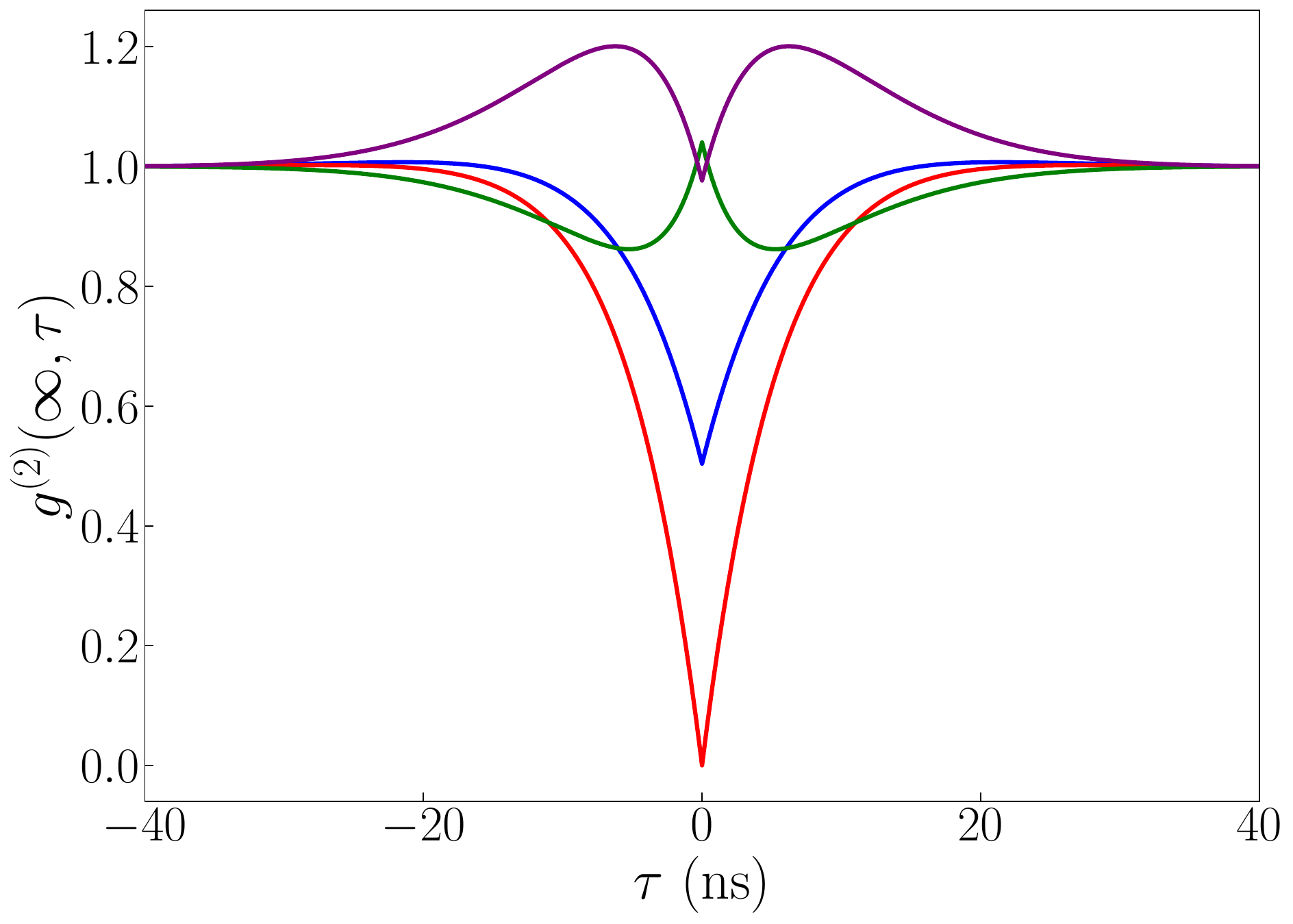} 
                \put(-5,65){
                \begin{tikzpicture}
                \node[anchor=north west] at (0, 100) {(c)};
                \end{tikzpicture}}
                \put(55,7){ 
                    \includegraphics[width=0.4\textwidth]{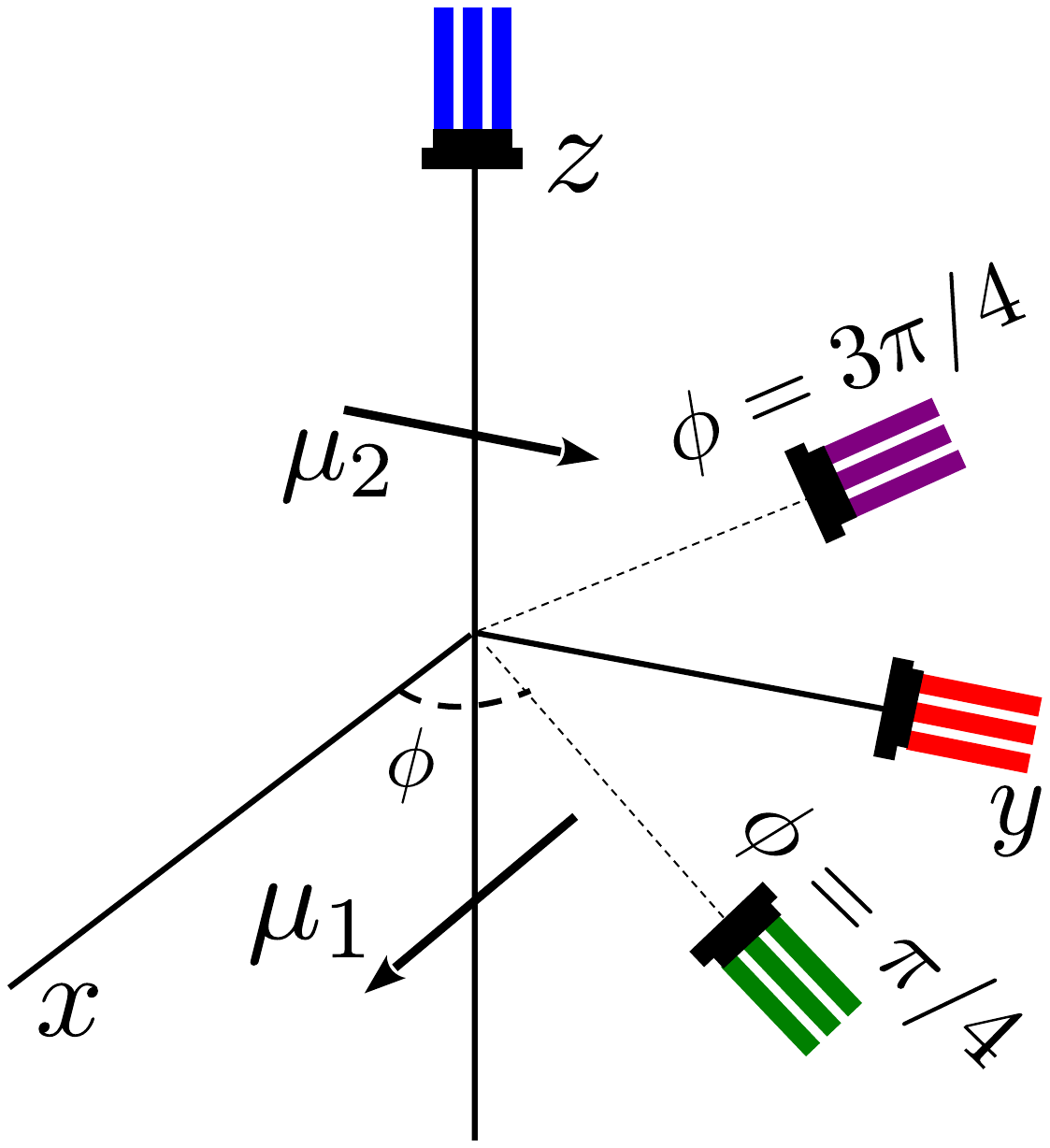}
                }
                \end{overpic} 
              \end{subfigure} 
              \caption{\justifying (a) Unnormalised second-order correlation $G^{(2)}(\infty, 0)$ and (b) normalised $g^{(2)}(\infty, 0)$ at zero time delay for an orthogonal dimer plotted as a function of the azimuthal angle $\phi$ (blue) and the spherical angle $\theta$ (red) in a polar plot. All angles are in radians. (c) Photon coincidence $g^{(2)}(\infty, \tau)$ plotted as a function of time-delay for different detection directions as indicated by the detector colour in the inset.} 
              \label{fig:dir_ortho} 
            \end{figure}
            The presence of quantum correlations in the photons emitted by orthogonal emitters depends on the direction along which we collect the two subsequent photons. 
            In Fig.~\ref{fig:dir_ortho}, we consider a limiting case and show four scenarios that may arise on sampling both photons along certain common detector directions, i.e. $\boldsymbol{q}=\boldsymbol{q}^{\prime}$.  
            The polar plots in Fig.~\ref{fig:dir_ortho}(a) and (b) show the unnormalised and normalised zero-delay photon coincidence, where, in both cases, its value goes 0.5 when both photons are detected perpendicular to the dipoles.
            However, the anti-dips with $g^{(2)}(\infty, 0) \approx 1$ along certain sampling directions is purely a result of the normalisation which we explain further below.
            The four distinct regimes which we observe in Fig.~\ref{fig:dir_ortho}(c), thus depend on the different states the system is projected into after detecting the first and second photon, giving rise to the so-called measurement-induced cooperativity \cite{PhysRevA.107.023718}. 
            Collecting the first photon with wave-vector $\boldsymbol{q} = (1, \pm 1, 0)$ (shown in Fig.~\ref{fig:dir_ortho}(c) with green and purple detector) leads to the preparation of a maximally entangled dark or bright state, respectively.
            The photon coincidence, then, takes an analytical form, $n_{ee}/(n_{\text{ee}} + n^{\boldsymbol{q}}_{A(S)})^{2}$. 
            In both cases, at steady-state, the system eigenstates are differently filled, with the bright state getting preferentially filled, because of incoherent pumping. This leads to the value of $g^{(2)}(0) > 1$ for the former and $g^{(2)}(0) < 1$ for the latter case. 
            As shown in Fig.~\ref{fig:dir_ortho}(c), the photon coincidence increases for measurement along $\boldsymbol{q} = (1, -1, 0)$ (purple detector) and decreases for that along $\boldsymbol{q} = (1, 1, 0)$ (green detector), at $\tau > 0$.
            This can be verified from the analytical expressions by calculating the first order derivative of $g^{(2)}$ at $\tau = 0$ (see Appendix~\ref{sec:analytical_g2}), hence determining whether it is an increasing or a decreasing function of time $\tau$.
            If the detectors now point along the direction of any one of the dipoles, they receive photons emitted only by the other dipole.
            The photon coincidence then dips to zero, as shown in Fig.~\ref{fig:dir_ortho}(c) (red curve), which is characteristic of a single emitter. 
            Upon positioning the detectors perpendicular to both the dipoles (along the direction of maximum emission), the zero-delay photon coincidence takes an analytical form $2n_{ee}/\left(2n_{\text{ee}} + n^{\boldsymbol{q}}_{S} + n^{\boldsymbol{q}}_{A}\right)^{2}$ [calculated using Eq.~\eqref{eqn:g2} and~\eqref{eqn:G2}]. 
            The blue curve in Fig.~\ref{fig:dir_ortho}(c), shows the signature of independent emitters, where $g^{(2)}(\infty, 0)$ goes to 0.5. This is because, at steady-state, the system eigenstates are equally populated and the intermediate $|\psi^{i = (g, e)}_{\boldsymbol{q}, \lambda} \rangle$ state separates into contributions from two independent channels each composed of the emitter eigenstates.  

\begin{figure}
                \centering
                \begin{overpic}[width=0.48\textwidth]{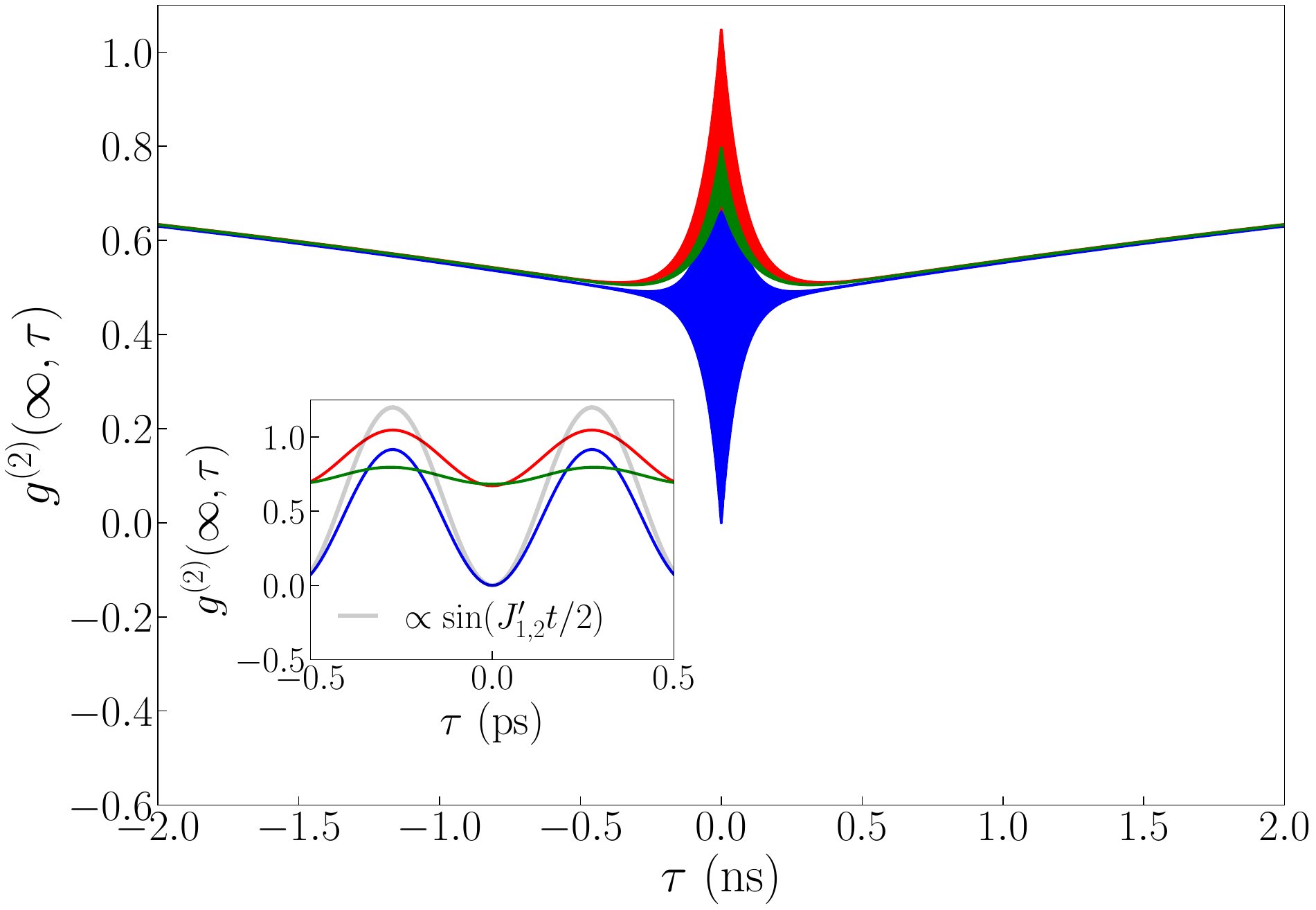} 
                \put(-2,63){
                \begin{tikzpicture}
                \node[anchor=north west] at (0, 100) {(a)};
                \end{tikzpicture}}
                \put(60,5.5){ 
                    \includegraphics[width=0.155\textwidth]{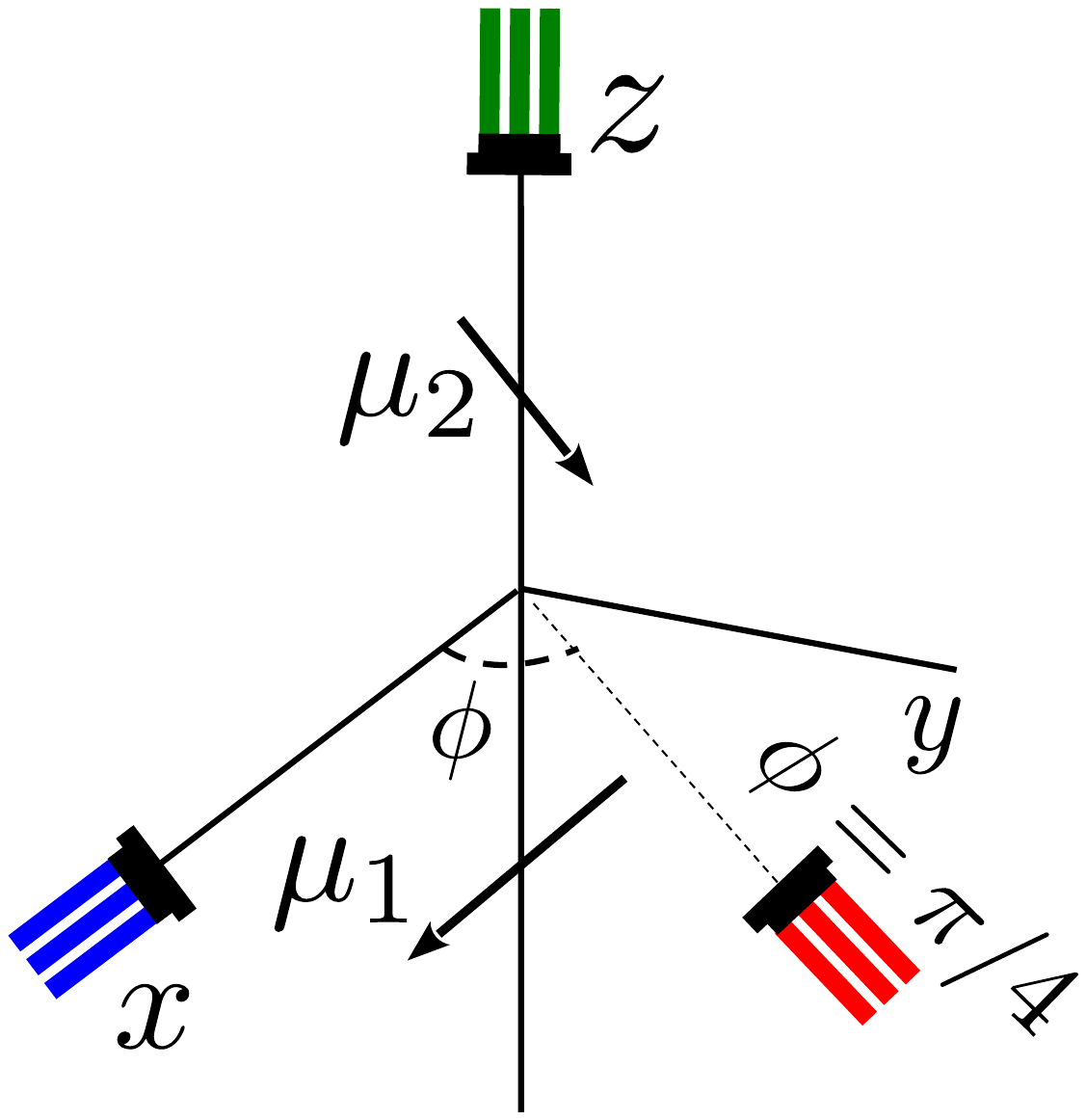}
                }
                \end{overpic} 

                \begin{overpic}[width=0.48\textwidth]{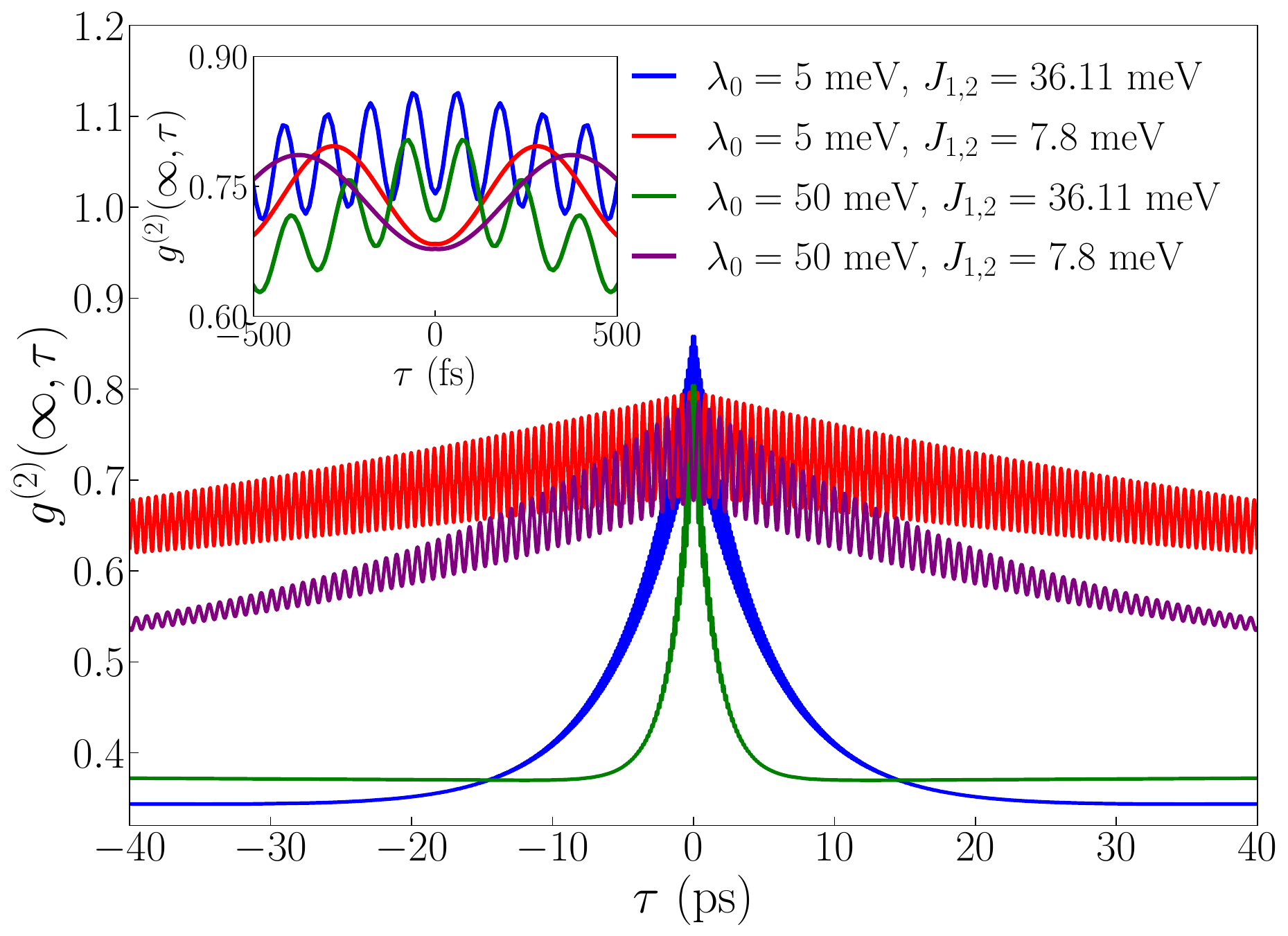} 
                \put(-2,66){
                \begin{tikzpicture}
                \node[anchor=north west] at (0, 100) {(b)};
                \end{tikzpicture}}
                \end{overpic} 
                \caption{\justifying Photon coincidence $g^{(2)}(\infty, \tau)$ for a $45\degree$ dimer.
                (a) Dependence on detection direction. The inset highlights coherent oscillations around $\tau \approx 0$, occurring at a frequency set by the renormalized dipole coupling $J_{1,2}^{\prime}$ (gray curve; see Appendix~\ref{sec:polaron_brme}).
                (b) Dependence on vibrational reorganization energy $\lambda$. Increasing $\lambda$ leads to faster dephasing of coherent oscillations and a narrowing of the $g^{(2)}$ envelope around zero delay, reflecting a reduced electronic coherence lifetime. The dipole-dipole coupling strength of $7.8$~meV and $36.11$~meV in (b) correspond to dimer separation of $2$~nm and $1.2$~nm, respectively.
                }
                \label{fig:dir_45}
            \end{figure}

            To further understand the role of dipole interactions, we now explore an intermediate configuration between the parallel and orthogonal dimers, where the monomers are aligned at $45\degree$ to each other.
            Just like the orthogonal case, projective measurements allow us to access any basis states, including the site basis. 
            However, the key difference is that, a finite dipole-dipole coupling in the $45\degree$ dimer introduces new dynamics in the photon coincidences in the form of coherent oscillations in $g^{(2)}(\infty, \tau)$ around zero time delay as shown in Fig.~\ref{fig:dir_45}.
            
            In Sec.~\ref{subsec:int}, we derived an expression for the intermediate state $|\psi^{i = (g, e)}_{\boldsymbol{q}, \lambda} \rangle$ into which the system is projected, depending on the photon detection direction.
            Now, for certain dimer configurations, such a projective measurement leads to the preparation of a state that is not an eigenstate of the system. 
            This leads to coherent oscillations on time scales set by the dipole coupling, arising from excitons oscillating between the sites \cite{doi:10.1021/acs.jpclett.1c01303, nation2024two}.
            These transient oscillations then decay at the phonon-induced decay rate in the long-time limit.

            When considering photons sampled perpendicular to both dipoles, i.e., along the $\boldsymbol{z}$-direction, we find that the height of the anti-dip decreases compared to Fig.~\ref{fig:orient_g2}, indicating a gradual transition from cooperative to independent emission upon changing the relative orientation from $0$ to $\pi/2$.
            When photons are measured perpendicular to either one (shown with red detector) or both dipoles (shown with green detector in Fig.~\ref{fig:dir_45}), the coherent oscillations occur with a reduced amplitude, albeit at the same frequency determined by the dipole coupling.
            This is because the intermediate states corresponding to an $\boldsymbol{x}$-polarised photon emitted ($|\psi^{i = (g, e)}_{\boldsymbol{q}, x} \rangle$) are coherent superposition states, while the ones corresponding to the $\boldsymbol{y}$-polarised photon are site basis states.
            This results in a suppressed oscillation amplitude when the contributions from both polarisation components are combined.
            Now, measuring photons along the dipole directions leads to detection of photons emitted only by the other dipole and we expect the $g^{(2)}(\infty, 0)$ to dip to zero as observed for orthogonal dimers.
            However, a non-zero dipole interaction gives rise to coherent oscillations, due to the preparation of site basis states $(\ket{e_{1}g_{2}}$ or $\ket{e_{2}g_{1}})$, depending on measurement direction. 

            The coherent oscillations we observe in the $g^{(2)}(\infty, \tau)$ occur on timescales determined by the renormalised dipole-dipole coupling and persist within the electronic coherence lifetime. 
            In our model, the vibrational reorganisation energy is assumed to be ($5$ meV \cite{B704962E}), which is smaller than the system-bath interactions present in many other dimeric systems studied using ultrafast spectroscopy \cite{sung2015direct, tempelaar2014mapping}. 
            Increasing the phonon reorganisation energy upto $50$~meV along with higher dipole-dipole interaction leads to more rapid dephasing of coherent oscillations. 
            Assuming perpendicular photon detection along $\boldsymbol{z}$ direction, we show in Fig. \ref{fig:dir_45}(b) that under these conditions the envelope of the photon coincidence signal shrinks markedly around zero-time delay and the electronic coherence lifetime is reduced to tens of picoseconds \cite{singh2021coherent}. 
            The corresponding absorption spectra of the $45\degree$ dimer, presented in Appendix~\ref{app:absorption}, provide further evidence for this behaviour, showing how increasing $\lambda_{0}$ both renormalises the excitonic splitting and accelerates optical dephasing, fully consistent with the observed narrowing of the photon coincidence envelope.

        
    \section{Factors affecting zero-delay coincidence}  
    \label{sec:factors}
        In this section, we focus on modelling different scenarios which might arise in an actual experiment ranging from modulating phonon bath temperature to effects of different degrees of orientational disorder.
        We calculate $g^{(2)}(\infty, 0)$ for two distinct configurations, H- and J-dimers and find that the dark state accessibility plays a major role in determining the height of the anti-dip, which is a signature of interemitter coherence \cite{PhysRevA.107.023718}.
    
    \subsection{Effect of vibrational bath temperature}
    \label{subsec:temp}
        In many experiments, lowering the sample temperature is key to stabilizing quantum emitters and minimizing environmental noise.
        In solid-state platforms, such as quantum dots and colour centers, cooling to cryogenic temperatures is essential for reducing phonon-induced decoherence \cite{Warburton2013}.
        Similarly, in biological light-harvesting complexes, temperature-dependent experiments have provided insight into coherent transfer \cite{engel2007evidence, Collini2010,KIM20191918}.
        Here, we explore the role of temperature in cooperative signatures by calculating the zero-delay coincidence across a range of vibrational bath temperature, from absolute zero to room temperature.
        Lowering the temperature suppresses thermal fluctuations, potentially enhancing and modifying the accessibility of dark states.
        \begin{figure}
            \centering
            \includegraphics[width=0.48\textwidth]{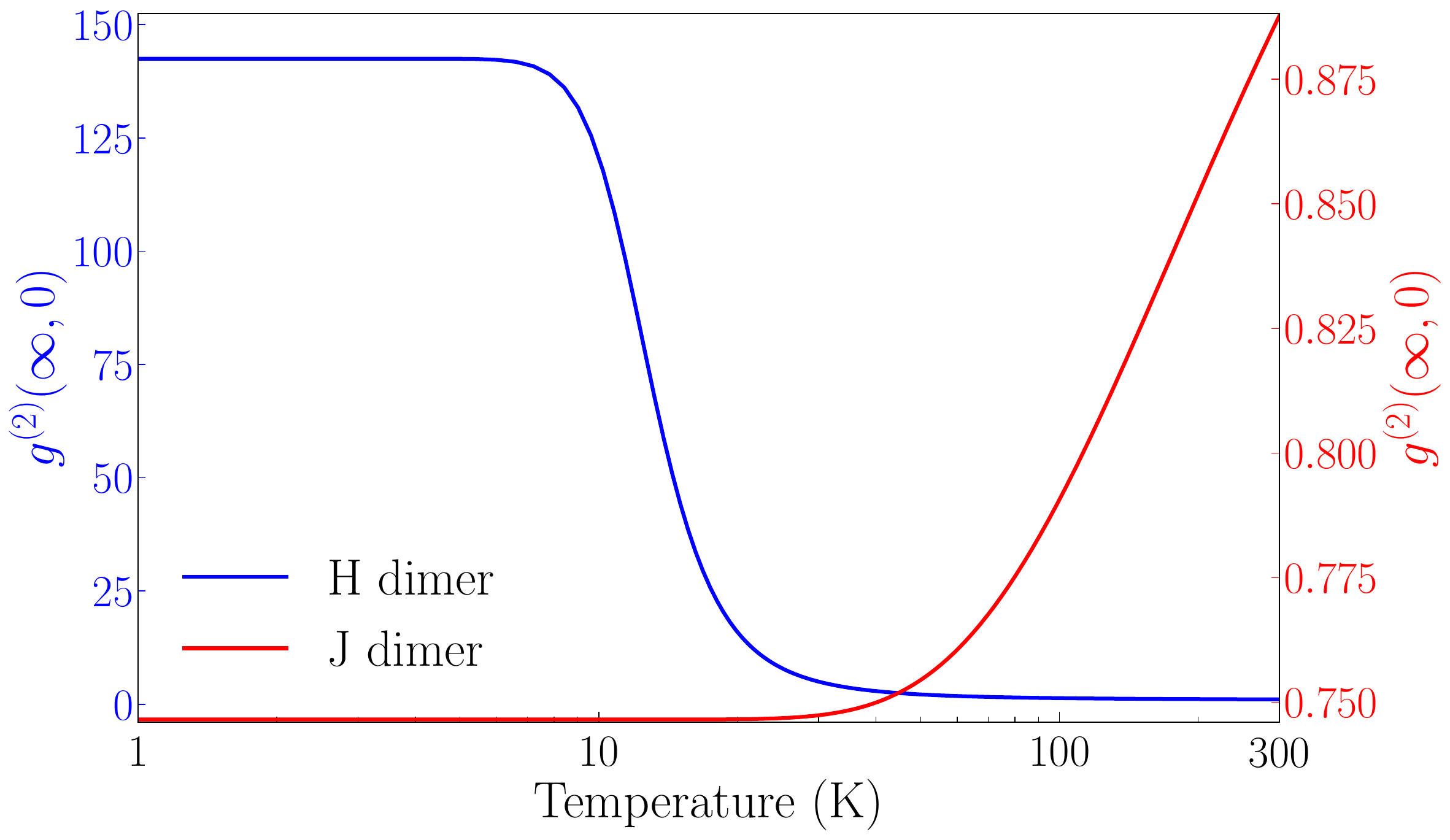}
            \caption{\justifying
            Zero-delay $ g^{(2)}(\infty, 0) $ as a function of phonon bath temperature. The left vertical axis (blue) corresponds to the H-dimer, and the right vertical axis (red) to the J-dimer. The magnitude of dipole–dipole coupling strength $J_{1, 2}$ is $ 7.8~\mathrm{meV} $ for the H-dimer and $ 15.6~\mathrm{meV} $ for the J-dimer (determined using Eq. \eqref{eqn:forst_coup}). Dipole parameters used: transition energy $\omega_{s} = 1.8~\mathrm{eV}$, separation distance $r_{12} = 2~\mathrm{nm}$, and dipole moment $|\boldsymbol{\mu}| = 10~\mathrm{D}$ for both emitters.).
            }
        \label{fig:g2_TempDep}
        \end{figure}
        Fig.~\ref{fig:g2_TempDep} shows the temperature dependence of $g^{(2)}(\infty, 0)$ for the H- and J-dimers.
        The cascade of transitions through the bright state in the H-dimer ends up trapped in the dark state at lower temperatures.
        As a result, fewer excitations decay radiatively to the ground state and are consequently less likely to be detected.
        As we found in Sec.~\ref{subsec:ph_coin}, the value of the normalised coincidence at zero-time delay is inversely proportional to the square of the bright state population.
        This normalisation essentially causes the value of $g^{(2)}(\infty, 0)$ to go up at very low temperatures.
        However, phonon assisted re-excitation back into the bright state at higher temperatures causes this value to plummet, as shown in Fig.~\ref{fig:g2_TempDep}.
        A J-dimer, on the other hand, does not show such a drastic change owing to the relative inaccessibility of the dark state even at higher temperatures causing the height of the anti-dip to only slightly change with rising temperature.
        
    \subsection{Effects of Ensemble Averaging}
    \label{subsec:ens_avg}   
\begin{figure*}[htbp]
        \centering
        \begin{subfigure}[b]{\textwidth}
            \centering
            \begin{overpic}[width=\textwidth]{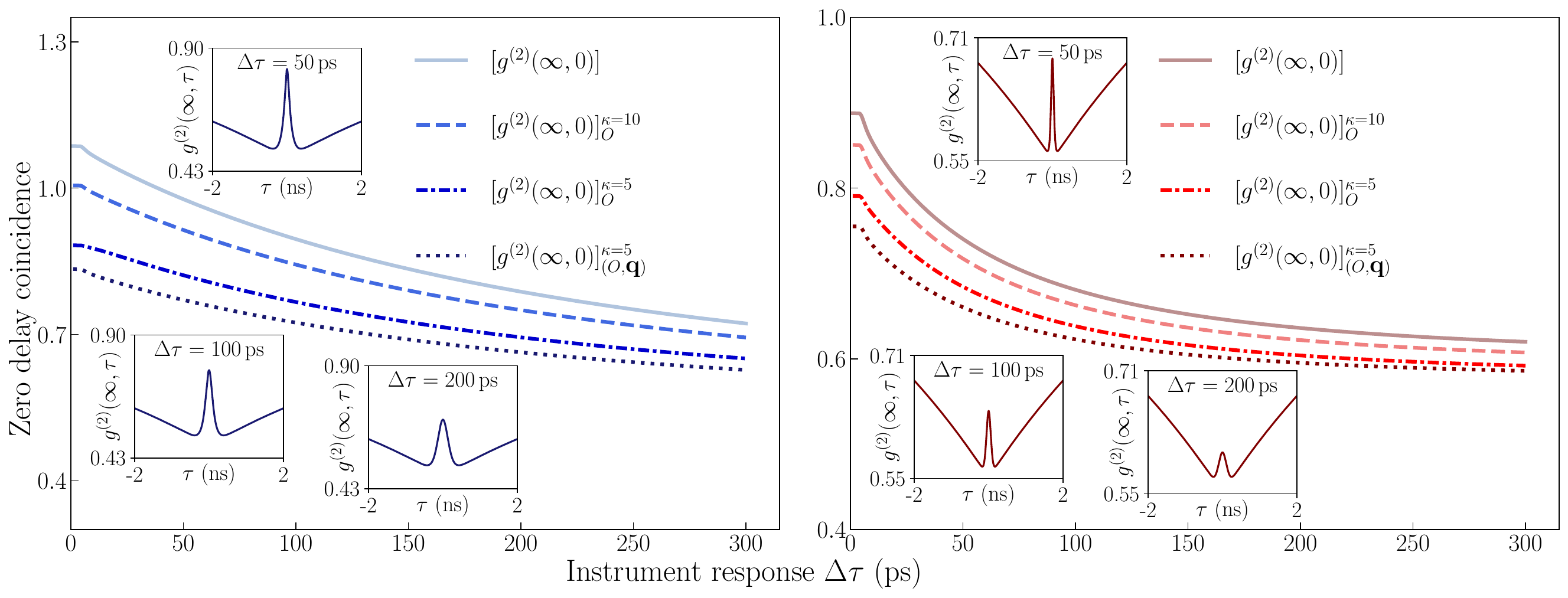}
                \put(-0.9,34){\begin{tikzpicture}[scale=0.2]
                \node[anchor=north west] {(a)};
                \end{tikzpicture}}
                \put(12,18.3){
    \begin{tikzpicture}[scale=0.2]
        \draw[draw=blue] (0,0) circle (0.3cm); 
        
        \draw[dashed, gray, thin] (-0.3,0) -- (0.39,6.6); 
        \draw[dashed, gray, thin] (0.3,0) -- (9,7.5);  
    \end{tikzpicture}
}

                \put(42,33){
                    \begin{tikzpicture}[scale=0.2]
                        \draw[thick,->,>=stealth,black] (-0.5,-1.0) -- (-0.5,1.5);
                        \draw[thick,->,>=stealth,black] (1,-1.0) -- (1,1.5);
                        \shade[ball color=blue] (-0.5,0) circle (0.5cm);
                        \shade[ball color=blue] (1,0) circle (0.5cm);
                        \begin{scope}[xshift=5cm, rotate=270]
                        \draw[very thick,gray] (0,0) -- (0,-1.5);
                        \draw[very thick,gray] (-0.3,0) -- (-0.3,-1.5);
                        \draw[very thick,gray] (0.3,0) -- (0.3,-1.5);
                        \fill[gray] (-0.3,-1.5) rectangle (0.3,-2);
                        \fill[black] (-0.4,-2) rectangle (0.4,-2.5);
                    \end{scope}
                    \end{tikzpicture}
                }
                \put(10.5,9.2){ 
    \begin{tikzpicture}[scale=0.2]
        \draw[draw=blue] (0,0) circle (0.3cm); 
        
        \draw[dashed, gray, thin] (-0.3,0) -- (-8.2,-0.4); 
        \draw[dashed, gray, thin] (0.3,0) -- (-1.6,-7.2);  
    \end{tikzpicture}
}

\put(42,26){
    \begin{tikzpicture}[scale=0.2]
        \draw[thick,->,>=stealth,black] (-0.5,-1.0) -- (-0.5,1.5);
        \draw[thick,->,>=stealth,black] (0.5,-1) -- (1.8,1.3);
        \shade[ball color=blue] (-0.5,0) circle (0.5cm);
        \shade[ball color=blue] (1,0) circle (0.5cm);
        \begin{scope}[xshift=5cm, rotate=270]
            \draw[very thick,gray] (0,0) -- (0,-1.5);
            \draw[very thick,gray] (-0.3,0) -- (-0.3,-1.5);
            \draw[very thick,gray] (0.3,0) -- (0.3,-1.5);
            \fill[gray] (-0.3,-1.5) rectangle (0.3,-2);
            \fill[black] (-0.4,-2) rectangle (0.4,-2.5);
        \end{scope}
    \end{tikzpicture}
}
\put(25,8.2){
    \begin{tikzpicture}[scale=0.2]
        \draw[draw=blue] (0,0) circle (0.3cm); 
        
        \draw[dashed, gray, thin] (-0.3,0) -- (-7.6,-0.6); 
        \draw[dashed, gray, thin] (0.3,0) -- (-0.7,-6.5);  
    \end{tikzpicture}
}
\put(40, 24){\begin{tikzpicture}
        \node at (-3, 0) {\scalebox{1.5}{$\bigg\}$}};
    \end{tikzpicture}
    }
                
                \put(42,18){
                    \begin{tikzpicture}[scale=0.2]
                    
                    \draw[thick,->,>=stealth,black] (-0.5,-1.0) -- (-0.5,1.5);
                    
                    \draw[thick,->,>=stealth,black] (0.5,-1) -- (1.8,1.3);
                    
                    \shade[ball color=blue] (-0.5,0) circle (0.5cm);
                    
                    \shade[ball color=blue] (1,0) circle (0.5cm);
                
                    \draw[dotted,thick] 
                    (-1,-2) 
                    arc[start angle=240, end angle=360, radius=3];

                    \begin{scope}[xshift=4cm, yshift=-2.5cm, rotate=235] 
                        \draw[very thick,gray] (0,0) -- (0,-1.5);
                        \draw[very thick,gray] (-0.3,0) -- (-0.3,-1.5);
                        \draw[very thick,gray] (0.3,0) -- (0.3,-1.5);
                        \fill[gray] (-0.3,-1.5) rectangle (0.3,-2);
                        \fill[black] (-0.4,-2) rectangle (0.4,-2.5);
                    \end{scope}
                \end{tikzpicture}
                }
            \put(49.5,34){\begin{tikzpicture}[scale=0.2]
                \node[anchor=north west] {(b)};
                                
                \end{tikzpicture}}

                \put(61.5,17.5){ 
    \begin{tikzpicture}[scale=0.2]
        \draw[draw=red] (0,0) circle (0.3cm);

        \draw[dashed, gray, thin] (-0.3,0) -- (-0.7,7.5); 
        \draw[dashed, gray, thin] (0.3,0) -- (7.5,7.3);  
    \end{tikzpicture}
}
                    \put(90,31.7){
                    \begin{tikzpicture}[scale=0.19]
                        \draw[thick,->,>=stealth,black] (0,-0.5) -- (0,2);
                        \draw[thick,->,>=stealth,black] (0,-2.5) -- (0,-0.2);
                        \shade[ball color=red] (0,0.5) circle (0.5cm);
                        \shade[ball color=red] (0,-1.5) circle (0.5cm);
                        \begin{scope}[xshift=4cm, yshift=0cm, rotate=270]
                        \draw[very thick,gray] (0,0) -- (0,-1.5);
                        \draw[very thick,gray] (-0.3,0) -- (-0.3,-1.5);
                        \draw[very thick,gray] (0.3,0) -- (0.3,-1.5);
                        \fill[gray] (-0.3,-1.5) rectangle (0.3,-2);
                        \fill[black] (-0.4,-2) rectangle (0.4,-2.5);
                    \end{scope}
                    \end{tikzpicture}
                }
                                \put(58,8.4){ 
    \begin{tikzpicture}[scale=0.2]
        
        \draw[draw=red] (0,0) circle (0.3cm); 
        
        \draw[dashed, gray, thin] (-0.3,0) -- (-10,-0.8);
        \draw[dashed, gray, thin] (0.3,0) -- (-1.5,-7.1);  
    \end{tikzpicture}
}
                \put(90,26){
                    \begin{tikzpicture}[scale=0.2]
                        \draw[thick,->,>=stealth,black] (-0.7,0) -- (1,1.5);
                        
                        \draw[thick,->,>=stealth,black] (0,-2.5) -- (0,-0.2);
                        
                        \shade[ball color=red] (0,0.5) circle (0.5cm);
                        
                        \shade[ball color=red] (0,-1.5) circle (0.5cm);
                        \begin{scope}[xshift=3.8cm, yshift=-0.3cm, rotate=270]
                    
                        \draw[very thick,gray] (0,0) -- (0,-1.5);
                        \draw[very thick,gray] (-0.3,0) -- (-0.3,-1.5);
                        \draw[very thick,gray] (0.3,0) -- (0.3,-1.5);
                        
                        \fill[gray] (-0.3,-1.5) rectangle (0.3,-2);
                        
                        \fill[black] (-0.4,-2) rectangle (0.4,-2.5);
                    \end{scope}
                    \end{tikzpicture}
                }
                \put(88,24){\begin{tikzpicture}
                    \node at (-3, 0) {\scalebox{1.5}{$\bigg\}$}};
                \end{tikzpicture}
                }
                \put(74,6.5){ 
    \begin{tikzpicture}[scale=0.2]
        
        \draw[draw=red] (0,0) circle (0.3cm); 
        
        \draw[dashed, gray, thin] (-0.3,0) -- (-8,-0.3); 
        \draw[dashed, gray, thin] (0.3,0) -- (-0.6,-7.5);  
    \end{tikzpicture}
}
                \put(90,19){
                    \begin{tikzpicture}[scale=0.2]
                    \draw[thick,->,>=stealth,black] (-0.7,0) -- (1,1.5);
                    \draw[thick,->,>=stealth,black] (0,-2.5) -- (0,-0.2);
                    \shade[ball color=red] (0,0.5) circle (0.5cm);
                    \shade[ball color=red] (0,-1.5) circle (0.5cm);
                    \draw[dotted,thick] 
                    (-1,-3) 
                    arc[start angle=240, end angle=360, radius=3];

                    \begin{scope}[xshift=3cm, yshift=-3.5cm, rotate=235] 
                        \draw[very thick,gray] (0,0) -- (0,-1.5);
                        \draw[very thick,gray] (-0.3,0) -- (-0.3,-1.5);
                        \draw[very thick,gray] (0.3,0) -- (0.3,-1.5);
                       
                        \fill[gray] (-0.3,-1.5) rectangle (0.3,-2);
                        
                        \fill[black] (-0.4,-2) rectangle (0.4,-2.5);
                    \end{scope}
                \end{tikzpicture}
                }
            \end{overpic}
            \end{subfigure}
        
            \caption{\justifying Zero-delay photon coincidence $g^{(2)}(\infty, 0)$ plotted against the detector's instrument response for two different dimer configurations: (a) H-dimer and (b) J-dimer. 
            The different line styles correspond to: an ideal case with no disorder (solid line), orientation disorder only [dashed ($\kappa_{\text{orient}} = 10$)] and different degrees of orientation plus detection-mode disorder [dot-dashed ($\kappa_{\text{orient}} = 5$) and dotted line ($\kappa_{\text{orient}} = 5$)]. 
            Insets show the corresponding photon coincidence $g^{(2)}(\infty, 0)$ at selected instrument response widths ($\Delta\tau = 50$, 100, 200 ps), illustrating the effect of increasing temporal resolution. The spin-pair illustrations on the right depict ideal alignment (top), disordered orientations (middle), and detection-mode mismatch (bottom), highlighting the physical origin of the different cases.}
            \label{fig:zero_time_g2_H_J}
        \end{figure*}

        Crystallisation of proteins such as Green Fluorescent Proteins (GFPs) produces dimers with relatively well-defined dipole orientations and reduced spatial disorder \cite{REKAS200250573, Ahmed2023}. 
        While our system is not specific to GFP, such examples illustrate that crystallisation can restrict molecular motion and facilitate directional photon collection; however, some residual dipole orientation averaging may still be present, particularly in partially ordered systems or ensemble measurements.
        To model orientational averaging, we introduce static disorder in one of the transition dipole moments $\boldsymbol{\mu}_{m}$ of the dimer by sampling its orientation from a von Mises–Fisher distribution centered around the mean dipole direction $\boldsymbol{\mu}_0$. This distribution, defined on the unit sphere, is given by:
        \begin{equation} 
            \begin{aligned} 
                f(\boldsymbol{\mu}_m \mid \boldsymbol{\mu}_0, \kappa) = C_p(\kappa) \exp\left( \kappa \boldsymbol{\mu}_0 \cdot \boldsymbol{\mu}_m \right), \end{aligned} 
        \end{equation}
        where $\kappa$ is the concentration parameter that controls the spread of orientations around the mean direction—larger values of $\kappa$ correspond to stronger clustering. The normalization constant $C_p(\kappa)$ for a 3D unit vector (i.e., $p = 3$) is given by
        \begin{equation} 
            \begin{aligned} 
                C_3(\kappa) = \frac{\kappa}{4\pi \sinh \kappa}. 
            \end{aligned} 
        \end{equation}
        Fig.~\ref{fig:zero_time_g2_H_J} shows the dependence of the zero-delay photon coincidence on the instrument response of the photon detectors. 
        We also consider a control scenario where the system is ideal, with fixed dipole orientations and no disorder in either the transition dipole moment or the detection direction.
        The presence of orientational disorder is denoted by $[g^{(2)}(\infty, 0)]_{O}$, whereas the static case without any disorder is denoted by $[g^{(2)}(\infty, 0)]$.

        Assuming a concentration parameter $\kappa_{\text{orient}} = 10$, in Fig.~\ref{fig:zero_time_g2_H_J}(a) and (b), we introduce a small degree of disorder in the relative dipole orientation, modelling crystallised H- and J-dimers respectively.
        For reference, this corresponds to approximately $75$\% of dipoles deviating by less than $35\degree$ from the mean orientation.
        The zero-delay $g^{(2)}(\infty, 0)$ which defines the height of the anti-dip reduces with increasing instrument response time.
        The zero-delay value also reduces owing to averaging over various relative dimer orientations.
        As discussed in Sec.~\ref{subsec:ph_coin}, the intermediate regime seen in orthogonal dimers---where the zero-delay photon coincidence approaches 0.5---can influence the overall $g^{(2)}(\infty, 0)$ in the presence of orientational disorder.
        
        Most experiments, however, involve ensembles of molecular emitters with non-rigid configurations suspended in a solvent. 
        Recent works using single-molecule imaging \cite{Worthy2019, https://doi.org/10.1002/advs.202003167} have enabled the investigation of isolated dimer molecules, which may tumble in the solvent, to varying degrees depending on their molecular weight.
        We explore two different scenarios for a tumbling dimer:
        First, we can consider a heavier dimer with a large molecular weight, which tumbles slowly compared to the timescale of photon emission and detection.
        Secondly, we consider a lighter dimer that tumbles quite fast. 
        This causes the two photons to be emitted in two different directions.
        In addition to this, for dimers in a solvent, we model orientational disorder using a smaller concentration parameter $(\kappa_{\text{orient}} = 5)$, allowing for broader sampling of dipole orientations. 
        To simulate the effect of limited tumbling during emission, we sample the detection direction of the first photon randomly over a unit sphere, while the second photon detection direction is sampled from a von Mises-Fisher distribution with a higher concentration parameter $\kappa_{\text{detect}} = 50$, representing a more localised detection.
        In contrast, for the lighter dimer, we assume rapid tumbling between the emission of the two photons.
        Therefore, we sample both photon detection directions independently and uniformly over the unit sphere, reflecting greater angular variation due to faster tumbling. 
        In Fig.~\ref{fig:zero_time_g2_H_J}(a) and (b) (dot-dashed line), we see that increasing the orientational disorder (modelling a heavy dimer in a solvent), suppresses the anti-dip, indicating further loss of detectable coherent signatures.
        For the second situation involving a light dimer (denoted by $[g^{(2)}(\infty, 0)]_{O, \boldsymbol{q}}$ in Fig.~\ref{fig:zero_time_g2_H_J}), introducing disorder in both modes being collected causes an even more pronounced reduction in the anti-dip height relative to the previous scenarios.

        Taking the case of a lighter dimer, tumbling in a solvent, we look at the photon coincidences at various instrument response in the insets in Fig.~\ref{fig:zero_time_g2_H_J}(a) and (b).
        We see that despite the various sources of orientational and detection-direction disorder considered here, our results suggest a remarkable robustness of the photon correlation signals against such static averaging. 
        In fact, it is primarily the instrument response of the detectors that limits the visibility of quantum signatures such as the anti-dip in $g^{(2)}(\infty, 0)$. 
        Even under moderate levels of orientational and mode-collection disorder---representing realistic conditions in crystallised and freely tumbling dimers---cooperative effects remain discernible. 
        This points to the potential feasibility of observing such signatures in practical experimental setups, provided sufficient time resolution is available.

        The results presented here consider disorder in dimer orientation and detection direction, which partially emulate the effects of finite spatial and polarization resolution in realistic detectors. Although we do not explicitly model the full acceptance cones of experimental polarisers, our findings suggest that moderate uncertainties in detection direction and polarization do not completely destroy the coherent signatures in photon correlations. Extending the model to include finite mode and polarization resolution explicitly would be an interesting direction for future work and would allow a more quantitative assessment of robustness under experimentally relevant conditions.

    
    \section{Summary}
        \label{sec:conc}
        In this work, we have developed a general theoretical framework to study emission characteristics of molecular dimers, accounting for strong vibrational coupling and directional photon detection.
        Using a Bloch-Redfield master equation within the polaron framework, we modelled the influence of a vibrational environment on quantum coherence between emitters. 
        By solving the Heisenberg's equation of motion, we calculated mode-resolved intensities and photon correlations, demonstrating how cooperative effects depend on the dimer configuration, detection direction, and vibrational coupling.
        Our results revealed that the optical signals obtained from such molecular dimers are influenced by the position of the dark state in the distinct energy level structure of different dimer configurations.
        We also examined how quantum coherence can become manifest in photon coincidence measurements on orthogonal dimers.
        In such configurations, certain detection conditions can reveal cooperative features in the optical response, even though these coherences do not contribute to emission rate enhancements.
        We also found that a finite dipole coupling in a $45\degree$ dimer can give rise to coherent oscillations when projective measurement leads to the formation of a non-eigenstate.
        Furthermore, we investigated the impact of experimentally tunable parameters, including sample temperature, static disorder and ensemble averaging, on the resolvability of coherent signatures.
        Our findings suggest that by carefully controlling these factors, it is possible to enhance the robustness of the observed cooperative effects, making them more accessible in practical experiments.
    
        While transforming observables from the polaron frame to the lab frame can suppress coherence contributions to both the intensity and the photon correlations \cite{10.1063/5.0228779, PhysRevResearch.6.033231}, we have not included these corrections explicitly in our calculations. For the reorganisation energies considered in this work, the Debye–Waller factor remains close to unity, making the effect of such corrections minimal. A more complete treatment of their influence, especially on two-time correlations, could be incorporated into future studies.
        
        Our framework can also be used to study different regimes of coherent and incoherent energy transfer, with increasing separation in such a two-emitter system, under improved detector response times and lower static disorder.
        Another obvious extension would be to study larger complex molecular structures, finding optimal pumping directions to populate specific energy levels of a system.
        The general theory developed in this work can also be employed to readily find system parameters such as dipole coupling or orientation in a biomolecular system from data.


    \section*{Acknowledgement}
       The authors thank Youngchan Kim and Steven S. Vogel for helpful discussions.
       P. Banerjee, A. Burgess and E. M. Gauger thank the Leverhulme Trust for support through Grant No. RPG2022-335. 
       J. Wiercinski and E. M. Gauger acknowledges financial support from Horizon Europe Pathfinder Challenge 101161312 APACE through Innovate UK Grant No. 10120741.
       M. Cygorek is supported by the Return Program of the State of North Rhine-Westphalia.
       
	
    \section{Data Availability}
        The data that support the findings of this study are available from the corresponding author upon reasonable request.
    		

    \appendix

    \section{Dipole projections along polarisation directions}\label{sec:dip_proj}
        Assuming a photon of mode $\boldsymbol{q}$ is sampled by the point-sized detectors, as shown in Fig.~\ref{fig:gen_k}, we can write the wave-vector $\boldsymbol{q} = |\boldsymbol{q}| \left(\sin \theta \cos \phi, \sin \theta \sin \phi, \cos \theta \right)$ and the distance vector separating the two dipoles $\boldsymbol{r} = |\boldsymbol{r}| \hat{\boldsymbol{r}}$, where
        \label{chap:dip_proj}
        \begin{equation}
            \begin{aligned}
            \hat{\boldsymbol{r}} & = \boldsymbol{e}_{z}\\
                \boldsymbol{e}_{y} & = \frac{\boldsymbol{r} \times \boldsymbol{\mu}_{1}}{\sqrt{r^{2} \mu_{1}^{2} - (\boldsymbol{r} \cdot \boldsymbol{\mu}_{1})^{2}}} = \frac{\boldsymbol{r} \times \boldsymbol{\mu}_{1}}{\Lambda} \\
                \boldsymbol{e}_{x} & = \boldsymbol{e}_{y} \times \boldsymbol{e}_{z} = \frac{(\boldsymbol{r} \times \boldsymbol{\mu}_{1}) \times\boldsymbol{r}}{\Lambda} = \frac{1}{\Lambda}\left[r \boldsymbol{\mu}_{1} - \frac{(\boldsymbol{r} \cdot \boldsymbol{\mu}_{1}) \boldsymbol{r}}{r}\right]
            \end{aligned}
        \end{equation}
        We can rearrange terms in $\boldsymbol{e}_{y}$ to write $\frac{\Lambda}{r} \boldsymbol{e}_{y} = \hat{\boldsymbol{r}} \times \boldsymbol{\mu}_{1}$. Performing a cross product w.r.t $\boldsymbol{e}_{z}$ on both sides \cite{PhysRevA.97.013819}, 
        \begin{equation}
            \begin{aligned}
                \frac{\Lambda}{r} \boldsymbol{e}_{z} \times \boldsymbol{e}_{y} & = \boldsymbol{e}_{z} \times (\hat{\boldsymbol{r}} \times \boldsymbol{\mu}_{1}) \\
                \frac{\Lambda}{r} \boldsymbol{e}_{x} & = \boldsymbol{\mu}_{1} - (\boldsymbol{\mu}_{1} \cdot \hat{\boldsymbol{r}}) \boldsymbol{e}_{z}\\
                \boldsymbol{\mu}_{1} & = \frac{\Lambda}{r} \boldsymbol{e}_{x} + (\boldsymbol{\mu}_{1} \cdot \hat{\boldsymbol{r}}) \boldsymbol{e}_{z}
            \end{aligned}
        \end{equation}
        \begin{figure}[h]
        \centering
        \tdplotsetmaincoords{70}{110} 
    
        \begin{tikzpicture}[tdplot_main_coords]
    
            \draw[thick,-] (0,0,0) -- (3,0,0) node[anchor=north east]{$x$}; 
            \draw[thick,-] (0,0,0) -- (0,2,0) node[anchor=north west]{$y$}; 
            \draw[thick,-] (0,0,-2) -- (0,0,2) node[anchor=south]{$z$}; 
            
            \draw[thick,-Latex] (-0.5,0.2,1) -- (0.2,-0.6,1) node[anchor=south west,above=1.2mm]{$\boldsymbol{\mu}_{1}$};
            \draw[thick,-Latex] (-0.8,0,-1) -- (1,0,-1) node[anchor=north west,below=0.2mm]{$\boldsymbol{\mu}_{2}$};
            
            \tdplotsetcoord{P}{4}{40}{45} 
            
            \draw[thick,-Latex] (0,0,0) -- (P) node[anchor=south west, above=-0.9mm]{$\mathbf{q}(\theta, \phi)$};
            
            \coordinate (O) at (2, 1.42, 2.2); 
            
            \draw[thick,-Latex] (O) -- ++(-1.6, 0.5, -0.4) node[anchor=south west, right=1.2pt, above=0.2pt]{$\lambda_1$};
            \draw[thick,-Latex] (O) -- ++(1.0, 0.8, 0) node[anchor=south west, right=1.2pt,below=0.2pt]{$\lambda_2$};
    
        \end{tikzpicture}
        \caption{\justifying An arbitrary dipole configuration with the wave-vector $\mathbf{q}$ and polarisation directions $\lambda_1$ and $\lambda_2$.}
        \label{fig:gen_k}
        \end{figure}
        Now, $\boldsymbol{\mu}_{2}$ can be written as $\sum_{i} (\boldsymbol{\mu}_{2} \cdot \boldsymbol{e}_{i})\boldsymbol{e}_{i}$.
        \begin{equation}
            \begin{aligned}
                \boldsymbol{\mu}_{2} & = (\boldsymbol{\mu}_{2} \cdot \boldsymbol{e}_{x})\boldsymbol{e}_{x} + (\boldsymbol{\mu}_{2} \cdot \boldsymbol{e}_{y})\boldsymbol{e}_{y} + (\boldsymbol{\mu}_{2} \cdot \boldsymbol{e}_{z})\boldsymbol{e}_{z} \\
                & = \left(\boldsymbol{\mu}_{2} \cdot \left(\frac{1}{\Lambda}\left[r \boldsymbol{\mu}_{1} - \frac{(\boldsymbol{r} \cdot \boldsymbol{\mu}_{1}) \boldsymbol{r}}{\boldsymbol{r}} \right] \right) \right) \boldsymbol{e}_{x} \\
                & + \left(\boldsymbol{\mu}_{2} \cdot \left(\frac{\boldsymbol{r}}{\Lambda} \times \boldsymbol{\mu_{1}}\right)\right) \boldsymbol{e}_{y} + (\boldsymbol{\mu}_{2} \cdot \hat{\boldsymbol{r}}) \boldsymbol{e}_{z}
            \end{aligned}
        \end{equation}
        The unit vectors along the polarisation direction perpendicular to the wave vector $\boldsymbol{q}$,
        \begin{align}
        \boldsymbol{\lambda}_{1} &= \left[\cos \theta \cos \phi, \cos \theta \sin \phi, -\sin \theta \right], \label{eq:ek1} \\
        \boldsymbol{\lambda}_{2} &= [-\sin \phi, \cos \phi, 0]. \label{eq:ek2}
        \end{align}
        Now, we have reached a point where we can evaluate the projections,
        \begin{align}
        \boldsymbol{\mu}_{1} \cdot \boldsymbol{\lambda}_{1} &= \frac{\Lambda}{r} \cos \theta \cos \phi - (\boldsymbol{\mu}_{1} \cdot \hat{\boldsymbol{r}}) \sin \theta \label{eq:mu1_ek1} \\
        \boldsymbol{\mu}_{1} \cdot \boldsymbol{\lambda}_{2} &= -\frac{\Lambda}{r} \sin \phi, \label{eq:mu1_ek2} \\
        \begin{split}
        \boldsymbol{\mu}_{2} \cdot \boldsymbol{\lambda}_{1} &= \frac{r}{\Lambda} \left[\boldsymbol{\mu}_{1} \cdot  \boldsymbol{\mu}_{2} - (\boldsymbol{\mu}_{1} \cdot \hat{\boldsymbol{r}})(\boldsymbol{\mu}_{2} \cdot \hat{\boldsymbol{r}}) \right] \cos \theta \cos \phi \\
        &\quad +  \frac{r}{\Lambda}\left[\hat{\boldsymbol{r}}\cdot \left(\boldsymbol{\mu}_{1} \times \boldsymbol{\mu}_{2}\right)\right]\cos \theta \sin \phi - (\boldsymbol{\mu}_{2} \cdot \hat{\boldsymbol{r}}) \sin \theta,
        \end{split} \label{eq:mu2_ek1} \\
        \begin{split}
        \boldsymbol{\mu}_{2} \cdot \boldsymbol{\lambda}_{2} &= -\frac{r}{\Lambda} \left[\boldsymbol{\mu}_{1} \cdot  \boldsymbol{\mu}_{2} - (\boldsymbol{\mu}_{1} \cdot \hat{\boldsymbol{r}})(\boldsymbol{\mu}_{2} \cdot \hat{\boldsymbol{r}}) \right] \sin \phi \\
        &\quad + \frac{r}{\Lambda}\left[\hat{\boldsymbol{r}}\cdot \left(\boldsymbol{\mu}_{1} \times \boldsymbol{\mu}_{2}\right)\right]\cos \phi.
        \end{split} \label{eq:mu2_ek2}
        \end{align}

        \section{Derivation of mode-resolved intensity}\label{sec:dir_int}
        The intensity profile for two coupled quantum emitters can be measured using a point-like detector which picks up photons along a certain wave-vector $\boldsymbol{q}$ \cite{PhysRevA.107.023718}. 
        The intensity of the signal obtained by the detector can be given as $I_{\boldsymbol{q}} = \sum_{\lambda}\frac{1}{\Delta \tau_{M}}\langle a_{\boldsymbol{q}, \lambda}^{\dagger} a_{\boldsymbol{q}, \lambda} \rangle$, where $\Delta \tau_{M}$ is the characteristic times taken for each such measurement.
        Solving the Heisenberg's equation of motion, we map the environment operators $a_{\boldsymbol{q}\lambda}$ and $a_{\boldsymbol{q}\lambda}^{\dagger}$ in $H_{\text{I,opt}}$ in Eq.~\eqref{eqn:H_opt} to the system operators $\sigma^{\pm}_{\boldsymbol{q}, \lambda}$ under the Markov approximation for the emission dynamics.       
        Using Heisenberg's equation of motion for the creation and annihilation operators,
        \begin{equation}
            \begin{aligned}
                \frac{\partial}{\partial t} a^{(\dagger)}_{\boldsymbol{q}, \lambda} & = \frac{i}{\hbar} [H, a^{(\dagger)}_{\boldsymbol{q}, \lambda}],  \\
                & = \mathcal{N}_{\boldsymbol{q}, \lambda} g_{\boldsymbol{q}} \sigma_{\boldsymbol{q}, \lambda}^{\mp} \mp i \omega_{\boldsymbol{q}, \lambda} a^{(\dagger)}_{\boldsymbol{q}, \lambda}. \\
                a^{(\dagger)}_{\boldsymbol{q}, \lambda}(t) &= a^{(\dagger)}_{\boldsymbol{q}, \lambda}(0) e^{\mp i \omega_{\boldsymbol{q}} t} + \mathcal{N}_{\boldsymbol{q}, \lambda} g_{\boldsymbol{q}} \int_{0}^{t} dt' e^{\mp i \omega_{\boldsymbol{q}}(t - t')} \sigma_{\boldsymbol{q}, \lambda}^{\mp}(t').\\
            \end{aligned}
        \end{equation}
        Assuming the optical environmentto be in a vacuum state  at time $t = 0$, we can write \cite{Agarwal1974},
        \begin{multline}
            a_{\boldsymbol{q}, \lambda}^{\dagger}(t)a_{\boldsymbol{q}, \lambda}(t) = \mathcal{N}_{\boldsymbol{q}, \lambda}^2 g_{\boldsymbol{q}}^{2} \int_{0}^{t} dt'' e^{-i \omega_{\boldsymbol{q}}t''} \sigma_{\boldsymbol{q}, \lambda}^{+}(t'') \\
                \int_{0}^{t} dt' e^{i \omega_{\boldsymbol{q}}t'}\sigma_{\boldsymbol{q}, \lambda}^{-}(t') 
        \end{multline}
        Since, $t'$ and $t''$ are dummy variables, we can introduce $I_{\pm}(t) = \int_{0}^{t} dt' e^{\mp i \omega_{\boldsymbol{q}}(t' - t)} \sigma_{\boldsymbol{q}, \lambda}^{\pm}(t')$.
        Thus, the time derivative of $a_{\boldsymbol{q}, \lambda}^{\dagger}(t)a_{\boldsymbol{q}, \lambda}(t)$ takes the form,
        \begin{equation}\label{eqn:rate_of_change}
        \frac{\partial}{\partial t} a_{\boldsymbol{q}, \lambda}^{\dagger}(t) 
        a_{\boldsymbol{q}, \lambda}(t) = \mathcal{N}_{\boldsymbol{q}, \lambda}^2 g_{\boldsymbol{q}}^2 \left[ 
        \sigma_{\boldsymbol{q}, \lambda}^{+}(t) I_{-}(t) + I_{+}(t) \sigma_{\boldsymbol{q}, \lambda}^{-}(t) \right].
        \end{equation}
            Using harmonic decomposition \cite{ficek2005quantum, 10.1116/5.0157714}, the system operators $\sigma_{1(2)}^{\pm}(t^{\prime})$, constituting $\sigma_{\boldsymbol{q}, \lambda}^{\pm}$ in Eq.~\eqref{eqn:ralo_op} can be written in terms of the delocalised basis states. 
            Now, performing the integration over $t^{\prime}$ we get the time local lowering operator, 
            we can write the mode-resolved intensity of two coupled emitters as \cite{PhysRevA.107.023718}
            \begin{equation}\label{eqn:dir_int_app}
                I_{\boldsymbol{q}}(t) = \sum_{\lambda = 1}^{2} 2\pi g_{\boldsymbol{q}}^{2} \mathcal{N}_{\boldsymbol{q}, \lambda}^{2} \delta(\omega_{\boldsymbol{q}} - \omega) \langle \sigma_{\boldsymbol{q}, \lambda}^{+}(t) \sigma_{\boldsymbol{q}, \lambda}^{-}(t) \rangle,
            \end{equation}
            where we assume a weak dipole interaction, i.e. $\omega_{S} \gg J_{1, 2}$ in Eq.~\eqref{eqn:system_ham}.

   

    \section{Polaron Master Equation}
                \label{sec:polaron_brme}                
                The polaron transformation from the lab to the polaron frame is given by the unitary operator $U_{P} = e^{G}$, where $G = \sum_{m,\boldsymbol{k}} \ket{m}\bra{m} g_{\boldsymbol{k}}(b_{m, \boldsymbol{k}}^{\dagger} + b_{m, \boldsymbol{k}}) / \omega_{m,\boldsymbol{k}}$ for multiple sites, each strongly coupled to the vibrational bath. 
                This can be further decomposed into the dipole basis as 
                \begin{equation}
                        e^{\pm G} = \ket{0}\bra{0} + \sum_{m} B_{m}^{\pm} \ket{m} \bra{m},
                \end{equation}
                where $B_{m}^{\pm} = \exp[\pm \sum_{m} g_{\boldsymbol{k}}(b_{m, \boldsymbol{k}}^{\dagger} - b_{m, \boldsymbol{k}}) / \omega_{m,\boldsymbol{k}}]$.
                Now, transforming into the polaron frame (labelled with a prime) rescales the transition energies and coupling in the dimer Hamiltonian,
                \begin{equation} 
                    \label{eqn:ham_pol_system}
                    H_{S}^{\prime} = \omega_{S}^{\prime} \sum_{m=1}^{2} \sigma_{m}^{+} \sigma_{m}^{-} + \frac{J_{1,2}^{\prime}(\boldsymbol{r}_{1,2})}{2} (\sigma_{1}^{+} \sigma_{2}^{-} + \sigma_{1}^{-} \sigma_{2}^{+}),
                \end{equation}  
                Here, the frequency is shifted by  the reorganisation energy of the phonon environment as $\omega^{\prime}_{S} = \omega_{S} - \lambda$, where
                \begin{equation}\label{eq:ph_reorg}
                    \lambda = \int_{0}^{\infty} \frac{\mathcal{J}_{\text{vib}}(\omega)}{\omega} d\omega.
                \end{equation}
                The renormalised dipole dipole coupling is $J^{\prime}_{1,2} = \kappa_{0}^{2} J_{1,2}$ where $\kappa_{0}$ is the expectation value of $B^{\pm}$ in the continuum limit,
                \begin{equation}
                    \kappa_{0} = \langle B^{\pm} \rangle = e^{-\frac{1}{2} \phi(0)},
                \end{equation}
                and
                \begin{equation}\label{eq:PhononProp}
                    \phi (t) = \int_{0}^{\infty} d\omega \frac{\mathcal{J}_{\text{vib}}(\omega)}{\omega^{2}} \left[\cos (\omega t) \coth (\frac{\beta \omega}{2}) - i \sin (\omega t)  \right].
                \end{equation}
                The polaron transformation diagonalises the strong exciton-phonon interaction Hamiltonian, at the cost of introducing another phonon-dependent dipole interaction term which gives rise to a new polaron frame vibrational Bloch-Redfield dissipator,
                \begin{multline}
                    \label{eqn:pol_coup_dis}
                    \mathcal{D}^{\prime}_{\text{coup}} = \sum_{n,m} \Gamma^{\text{coup}}_{n,m}(\omega_{m}) [A_{m}(\omega_{m}) \rho_{S}(t) A^{\dagger}_{n}(\omega_{n}) \\- A^{\dagger}_{n}(\omega_{n})A_{m}(\omega_{m}) \rho_{S}(t)] + \text{h.c.}
                \end{multline}
                This results in the intra-manifold transitions facilitated by strongly coupled photons in the presence of a strong dipole interaction. 
                The bath correlation functions capture the environmental contribution to the interaction and are given by the rates $\Gamma_{nm}(\omega)$, where 
                \begin{equation}
                    \begin{aligned}
                        \Gamma_{n,m}(\omega) & = \int_{0}^{\infty} e^{i \omega s} \langle E_{n}^{\dagger}(t) E_{m}(t - s) \rangle,
                        \end{aligned}
                \end{equation}
                where $E(t)$ represent the environment operators in the interaction Hamiltonian. 
                The rate of dissipation $\Gamma^{\text{coup}}_{n,m}(\omega_{m})$ in Eq.~\eqref{eqn:pol_coup_dis} has contributions from the pairwise combinations of `raising' and `lowering' operators, now including the renormalisation due to the polaron transformation. 
                Depending on whether they pertain to the same TLS or not, there can be four possible combinations,
                \begin{equation}
                    \begin{aligned}
                        \langle \hat{B}^{\pm}_{m}(t)\hat{B}^{\pm}_{n}(0) \rangle & = \kappa_{0}^{2}e^{-\phi(t)}, \\
                        \langle \hat{B}^{\mp}_{m}(t)\hat{B}^{\pm}_{n}(0) \rangle & = \kappa_{0}^{2}e^{\phi(t)},  \\       \langle \hat{B}^{\pm}_{m}(t) \rangle  \langle \hat{B}^{\pm}_{n}(0) \rangle & = \kappa_{0}^{2}, \\
                        \langle \hat{B}^{\mp}_{m}(t)\rangle  \langle \hat{B}^{\pm}_{n}(0) \rangle & = \kappa_{0}^{2}.
                    \end{aligned}
                \end{equation}
                where the summation over $n$ and $m$ encapsulates all pairwise combination of processes for a dipole.  
                Throughout the paper, we use a super-Ohmic spectral density of the form,
                \begin{equation}\label{eqn:sup_ohm}
                    \begin{aligned}
                        \mathcal{J}_{\text{vib}}(\omega) & = \frac{\lambda_{0}}{2 \omega_{c}^{3}} \omega^{3} e^{-\frac{\omega}{{\omega_c}}},
                    \end{aligned}
                \end{equation}
                to calculate the phonon-induced decay rate \cite{C7CP06237K, doi:10.1021/acs.jpclett.9b01349, Rouse_2019, PRXEnergy.2.013002}. 
                Here, $\lambda_{0}$ is the reorgranisation energy and $\omega_{c}$ is the cut-off frequency.
                
                The optical interaction under the polaron transformation takes in some of the vibrational degrees of freedom and is proportional to $(B^{+}_{m} \sigma^{+}_{m} + B^{-}_{m} \sigma^{-}_{m})$. We project the system operators into the system eigenbasis and taking a list of all processes, in the Schr\"odinger picture, the second term in Eq.~\eqref{eqn:born_markov} reduces to the non-secular optical dissipator in the polaron frame,
                \begin{multline}
                    \label{eqn:pol_opt_dis}
                    \mathcal{D}^{\prime}_{\text{opt}} = P_{\text{vib}} \sum_{n,m}
                        \mathcal{F}(\Gamma^{\text{opt}}_{n,m}(\omega_{m}) [A_{m}(\omega_{m}) \rho_{S}(t) A^{\dagger}_{n}(\omega_{n}) \\- A^{\dagger}_{n}(\omega_{n})A_{m}(\omega_{m}) \rho_{S}(t)] + \text{h.c.}),
                \end{multline}
                where the pairwise combinations of raising and lowering operators $A_{\alpha}^{(\dagger)}$ are associated with frequency $\mp \omega_{\alpha}$.
                The raising and lowering operators for the optical dissipator are the $\sigma^{\pm}_{\alpha}$ operators in the diagonal basis.
                Every term in the above expression is weighted by a cross function $\mathcal{F} = \boldsymbol{\mu}_{1} \cdot \boldsymbol{\mu}_{2}$ \cite{Rouse_2019, ficek2005quantum}. 
                $\mathcal{D}^{\prime}_{\text{opt}}$ in Eq.~\eqref{eqn:pol_opt_dis} leads to the intermanifold transitions due to the optical interactions where the rates associated with the vibrational and optical processes automatically separate out due to the difference in timescales. 
                The value of $P_{\text{vib}}$ thus, depends on the nature of $A_{\alpha}$. 
                $P_{\text{vib}} = 1$, if the operators belong to the same system or if they are a combination of ladder operators. 
                $P_{\text{vib}} = B^{4}$ for equal operators or else equal to $B^{2}$ if they belong to different systems. 
                
                Assuming the environment to be in a thermal state, we find the rate for cartesian coordinates $i, j$, the rates
                \begin{equation}\label{eqn:rate}
                     \Gamma^{\text{opt}}_{n_{i},m_{j}}(\omega) = \delta_{ij}\left(\frac{1}{2} \gamma(\omega) + i S(\omega)\right).
                \end{equation}
                We can write the rate as $\gamma(\omega) =  \mathcal{J}(\omega) N(\omega)$, where $\mathcal{J}(\omega)$ is the spectral density and
                \begin{align}
                    N(\omega) =
                    \begin{cases} 
                        1 + n(\omega), & \omega \geq 0 \\ 
                        n(\omega), & \omega < 0
                    \end{cases} \label{eq:N_omega}
                \end{align}
                where the Bose-Einstein occupation of the photon modes $n(\omega) = 1 / (\exp[\beta \hbar \omega] - 1)$, with $\beta = 1 / k_{B}T$. 
                Here T is the temperature of the optical bath.
                For the optical dissipator, we assume a flat spectral density $\mathcal{J}_{\text{opt}}(\omega) = k_{\text{opt}}$, where $k_{\text{opt}}$ is the spontaneous decay rate of a single dipole with lifetime $\tau_{L}$, which the spontaneous emission rate                
                \begin{equation}
                \gamma(\omega) = \frac{4\omega^3 |\boldsymbol{\mu}|^2}{3\hbar c^3}(1 + n(\omega)) = \gamma_{\mathrm{opt}}(1 + n(\omega)),
                \end{equation}
                where \(\gamma_{\mathrm{opt}}\) is the zero-temperature spontaneous decay rate of a two-level system \cite{Rouse_2019, doi:10.1021/acs.jpclett.9b01349}.


\section{Contribution of high-frequency modes in the vibrational bath}

In many realistic molecular systems, the vibrational environment may contain both low-frequency (LF) modes that are coupled to the system dynamics and high-frequency (HF) modes which predominantly contribute to static energy renormalisation. The total spectral density can then be decomposed into LF and HF parts,
\begin{equation}
    \mathcal{J}(\omega) = \mathcal{J}_{L}(\omega) + \mathcal{J}_{H}(\omega),
\end{equation}
where $\mathcal{J}_{L}(\omega)$ describes modes within the energy scale of system dynamics, while $\mathcal{J}_{H}(\omega)$ represents high-frequency vibrations.

The HF modes mainly induce a static renormalisation of system parameters. This can be treated by performing a partial polaron transformation on the HF contribution, while leaving the LF modes untransformed. As a result, the system operators are renormalised, 
\begin{equation}
    \sigma_x \longrightarrow \kappa_{H} \sigma_x,
\end{equation}
and the bare transition energies are shifted by the associated HF reorganisation energy,
\begin{equation}
    \omega_0 \longrightarrow\  \omega_0 - \lambda_{H}.
\end{equation}
We model the HF spectral component by a sum of super-Ohmic Gaussian peaks, 
\begin{equation}
    \mathcal{J}_{H}(\omega) = \sum_{i} \alpha_{i}\,\omega^{3}\,\exp\bigg[-\frac{(\omega-\omega_{i})^{2}}{\gamma_{i}^{2}}\bigg],
\end{equation}
where each HF mode $i$ is centred at frequency $\omega_{i}$ with width $\gamma_{i}$ and coupling amplitude $\alpha_{i}$. The total reorganisation energy is then
\begin{align}
    \lambda_{H} &= \sum_{i} \lambda_{H_i},\\[4pt]
    \lambda_{H_i} &\sim\alpha_{i}\sqrt{\pi}\,\omega_{i}^{2}\gamma_{i},
\end{align}
and the corresponding renormalisation factor for the system couplings is 
\begin{align}
    \kappa_{H} &= \prod_{i} \kappa_{H_i},\\[4pt]
    \kappa_{H_i} &\sim \exp \Bigg[-\alpha_{i}\,\frac{\sqrt{\pi}}{2}\omega_{i}\gamma_{i}\Bigg] 
                  = \exp\Bigg[-\frac{\lambda_{H_i}}{2\omega_{i}}\Bigg].
\end{align}
Thus, given the reorganisation energy $\lambda_{H_i}$ of each high-frequency mode, the overall static renormalisation is fully determined by reorganisation energy $\lambda_{H_i}$ and the energy of the modes $\omega_{i}$.

    \section{Analytical calculation of mode-resolved \texorpdfstring{$d g^{(2)}(\infty, t)/dt\big|_{t \to 0}$}{dg2/dt at t=0 for infinity} for an Orthogonal dimer}

    \label{sec:analytical_g2}
In this section, we outline the derivation of the $\boldsymbol{q}$-dependent photon coincidence at zero time delay, $g^{(2)}_{\boldsymbol{q} = \boldsymbol{q}^{\prime}}(\infty, 0)$, for an orthogonal dimer, in terms of the ground state $(n_{\text{gg}})$, bright state $(n^{\boldsymbol{q}}_{\text{S}})$, dark state $(n^{\boldsymbol{q}}_{\text{A}})$, and doubly excited state $(n_{\text{ee}})$ population. 
The system is optically pumped with a rate $\gamma_p$, and the Bose-Einstein occupation number for the frequency associated with the transition, denoted by $n(\omega)$. 
Here, we consider two specific cases for orthogonal dimers, we explored in Sec.~\ref{subsec:ph_coin} and calculate the time-derivative of the photon coincidence at time-delay $\tau \to 0$. We can determine its behaviour by looking at the value of the first derivative $g^{(2)\prime}(\infty, 0)$.

For the measurement direction $\mathbf{q} = (-1, 1, 0)$, the time-derivative second-order correlation function at zero time delay can be calculated from the expression for $g^{(2)}(\infty, 0)$ in Sec.~\ref{subsec:ph_coin}:
\begin{equation}
    \begin{aligned}
        g^{(2)\prime}(\infty, 0) &= \frac{n^{\prime}_{\text{ee}}}{\big(n^{\boldsymbol{q}}_{\text{S}} + n_{\text{ee}}\big)^2} 
        - \frac{2 n_{\text{ee}} \big(n^{\boldsymbol{q}\prime}_{\text{S}} + n^{\prime}_{\text{ee}}\big)}{\big(n^{\boldsymbol{q}}_{\text{S}} + n_{\text{ee}}\big)^3},
    \end{aligned}
\end{equation}
where $n^{\prime}_{\text{ee}} = d n_{\text{ee}} / dt$ and $n^{\boldsymbol{q}\prime}_{\text{S}} = d n^{\boldsymbol{q}}_{\text{S}} / dt$.

The rate of change of the doubly-excited and symmetric state populations are given by:
\begin{equation}
    \begin{aligned}
        \frac{d n_{\text{ee}}}{dt} &= -2 \gamma \big(n(\omega) + 1\big) n_{\text{ee}} + \gamma n(\omega) \big(n^{\boldsymbol{q}}_{\text{S}} + n^{\boldsymbol{q}}_{\text{A}}\big) + \gamma_p n^{\boldsymbol{q}}_{\text{S}}, \\
        \frac{d n^{\boldsymbol{q}}_{\text{S}}}{dt} &= \gamma \big(n(\omega) + 1\big) n_{\text{ee}} - \gamma \big(2 n(\omega) + 1\big) n^{\boldsymbol{q}}_{\text{S}} \\
        &\quad + \gamma n(\omega) n_{\text{gg}} + \gamma_p \big(n_{\text{gg}} - n^{\boldsymbol{q}}_{\text{S}}\big).
    \end{aligned}
\end{equation}

For the measurement direction $\mathbf{q} = (1, 1, 0)$, the corresponding time-derivative of the correlation function is:
\begin{equation}
    \begin{aligned}
        g^{(2)\prime}(\infty, 0) &= \frac{n^{\prime}_{\text{ee}}}{\big(n^{\boldsymbol{q}}_{\text{A}} + n_{\text{ee}}\big)^2} 
        - \frac{2 n_{\text{ee}} \big(n^{\boldsymbol{q}\prime}_{\text{A}} + n^{\prime}_{\text{ee}}\big)}{\big(n^{\boldsymbol{q}}_{\text{A}} + n_{\text{ee}}\big)^3},
    \end{aligned}
\end{equation}
where rate equations for the populations are:
\begin{equation}
    \begin{aligned}
        \frac{d n_{\text{ee}}}{dt} &= -2 \gamma \big(n(\omega) + 1\big) n_{\text{ee}} + \gamma n(\omega) \big(n^{\boldsymbol{q}}_{\text{S}} + n^{\boldsymbol{q}}_{\text{A}}\big) \\
        &\quad + \gamma_p n^{\boldsymbol{q}}_{\text{S}},\\
        \frac{d n^{\boldsymbol{q}}_{\text{A}}}{dt} &= \gamma \big(n(\omega) + 1\big) n_{\text{ee}} - \gamma \big(2 n(\omega) + 1\big) n^{\boldsymbol{q}}_{\text{A}} \\
        &\quad + \gamma n(\omega) n_{\text{gg}}.
    \end{aligned}
\end{equation}

Note, that the pump term only appears in the case of detection along $\mathbf{q} = (1, 1, 0)$, since that projects the system into the symmetric state which is pumped. 

We can now use the values of the steady-state bright, dark and doubly excited state population, and optical decay and pumping rate to calculate the time-derivative of $g^{(2)}(\infty, \tau)$ at $\tau \to 0$.
Thus, the values of $g^{(2)\prime}(\infty, 0)$ will describe the photon correlation behavior at zero time delay for the orthogonal dimer configuration.
For detection along $\mathbf{q} = (-1, 1, 0)$ direction, we get a positive $g^{(2)\prime}(\infty, 0)$, which indicates a positive slope and hence an increasing photon correlation (hence a dip).
This explains the tall shoulders we see in the photon coincidence plotted in Fig.~\ref{fig:dir_ortho}.
In contrast, the negative value for $\mathbf{q} = (1, 1, 0)$ indicates a negative-slop of $g^{(2)}(\infty, 0)$ around zero-time delay, indicating an anti-dip.

\section{Absorption spectra of the $45\degree$ dimer}
\label{app:absorption}

\begin{figure}[h!]
\centering

\begin{overpic}[width=0.45\textwidth]{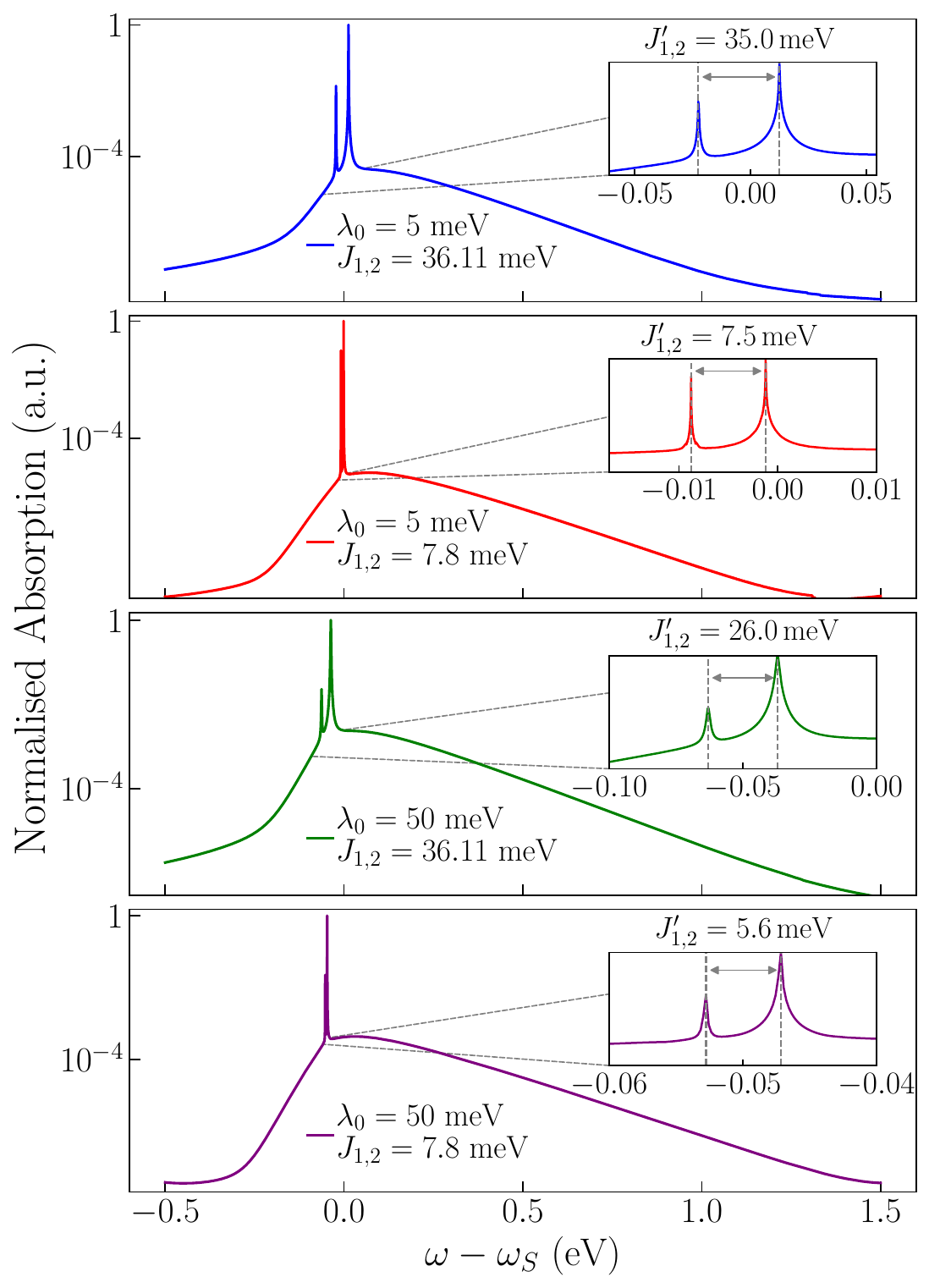}
    \put(5,95){(a)}
    \put(5,72){(b)}
    \put(5,49){(c)}
    \put(5,26){(d)}

\end{overpic}

\caption{\justifying
Normalised absorption spectra of the $45\degree$ dimer for different dipole-dipole couplings $J_{1,2}$ and vibrational reorganisation energies $\lambda_0$. 
Vertical dashed lines in the inset indicate the excitonic peak positions at $-\lambda_{0} \pm J_{1,2}^\prime / 2$, denoting the antisymmetric (right) and the symmetric state (left), respectively. The separation between these peaks gives the renormalised dipole coupling $J_{1,2}^\prime$. 
}
\label{fig:abs_45}
\end{figure}

To complement the discussion of photon coincidences and coherence lifetimes, we now consider the linear absorption spectra of the $45\degree$ dimer for different bath and dipole parameters, as shown in Fig.~\ref{fig:abs_45}. These spectra provide an intuitive picture of the energy landscape governing the electronic coherences.

Within the polaron master equation framework, the absorption spectrum is obtained from the Fourier transform of the two-time system correlation function at the steady state,
\begin{equation} 
S_{\mathrm{A}}(\omega) = \mathcal{R} \left\{\int_0^\infty d\tau e^{i\omega \tau} \sum_{m,n=1}^2 \gamma_{mn} \langle \sigma_m^-(t+\tau) \sigma_n^+(t) \rangle F_{mn}(\tau)\right\},
    \label{eq:abs_polaron}
\end{equation}
where, 
\begin{equation}
    F_{mn}(\tau) =
    \begin{cases}
    \kappa_0^2 e^{\phi(\tau)}, & m = n, \\
    \kappa_0^2 , & m \neq n.
    \end{cases}
\end{equation}
Here $\gamma_{mn}$ are the optical rates given in Eq.~\eqref{eqn:rate} and $\phi(\tau)$ is the phonon propagator introduced in Eq.~\eqref{eq:PhononProp}. 
The two-time expectation values are evaluated using the quantum regression theorem applied to the steady-state solution of the polaron master equation in Sec.~\ref{sec:polaron_brme}. 
The exponential factor $e^{\phi(\tau)}$ captures phonon-induced broadening. 
Consequently, the peak positions and linewidths directly reflect the renormalised transition energies $\omega_{S}^{\prime} = \omega_{S} - \lambda_{0}$ and coupling $J_{1,2}^{\prime} = \kappa_{0}^{2} J_{1,2}$, where $\kappa_{0} = e^{-\phi(0)/2}$.

For all configurations, the spectra exhibit two distinct peaks corresponding to the delocalised exciton states formed by coherent coupling between the monomers. The peak positions reflect the combined effect of the bare dipole-dipole interaction and renormalisation by vibrational dressing. The energy splitting between the peaks directly gives the renormalised dipole coupling $J_{1,2}^{\prime}$, which sets the timescale of coherent oscillations in $g^{(2)}(\infty,\tau)$ (see Fig.~\ref{fig:dir_45}). By changing the dipole separation from $2$~nm to $1.2$~nm we see the effect of weak and strong dipole couplings. For $\lambda_{0} = 5$~meV we find $J_{1,2}^{\prime} = 35$~meV (strong dipole coupling) and $7.5$~meV (weak dipole coupling), while for $\lambda_{0} = 50$~meV the splittings reduce to $26$~meV and $5.6$~meV, respectively.

In addition to renormalising the coupling, the phonon environment shifts the exciton resonance frequency by the phonon reorganisation energy $\lambda_{0}$ as calculated in Eq. \eqref{eq:ph_reorg}. Since we plot the spectra against $\omega - \omega_{S}$, the polaron shift can be inferred to directly from the peak positions [$\lambda_{0} = (\omega_{S} \pm J^{\prime}_{1,2} / 2) - \omega$] of the bright and dark states, respectively. We also find that increasing the phonon reorganisation energy accelerates dephasing, with broadening of the absorption peaks and a concomitant reduction in coherence lifetime. Finally, the polaron-modified coupling factor $\kappa_0 = \sqrt{J^{\prime}_{1, 2} / J_{1,2}}$ quantifies the reduction of the bare dipole interaction due to phonon dressing, providing a simple measure for the suppression of coherent oscillations in the system. We find the calculated values of $\kappa_0 \approx 0.98$ for $\lambda_0 = 5$~meV and $\kappa_0 \approx 0.85$ for $\lambda_0 = 50$~meV, in excellent agreement with the absorption spectra.
In practice, however, the extraction of the renormalised parameters requires prior knowledge of the underlying bare Hamiltonian parameters. Experimentally, one could distinguish these contributions by varying the temperature of the vibrational bath to partially suppress the phonon dressing - thereby reducing the reorganisation energy and recovering the bare excitonic splitting in the low-temperature limit. 


\section{Additional factors affecting zero-delay coincidence}
We now examine how site-level non-radiative decay and exciton–exciton annihilation (EEA) influence the zero-delay photon coincidence $g^{(2)}(\infty, 0)$. Both mechanisms introduce additional loss channels that could, in principle, alter the population dynamics of the intermediate states relevant for the two-photon cascades. For convenience we choose the direction of photon detection to be perpendicular to the dipole orientation and look chiefly at H- and J-dimers. 

\subsection{Site-level non-radiative decay}
\begin{figure}[htbp]
    \centering
    \includegraphics[width=0.45\textwidth]{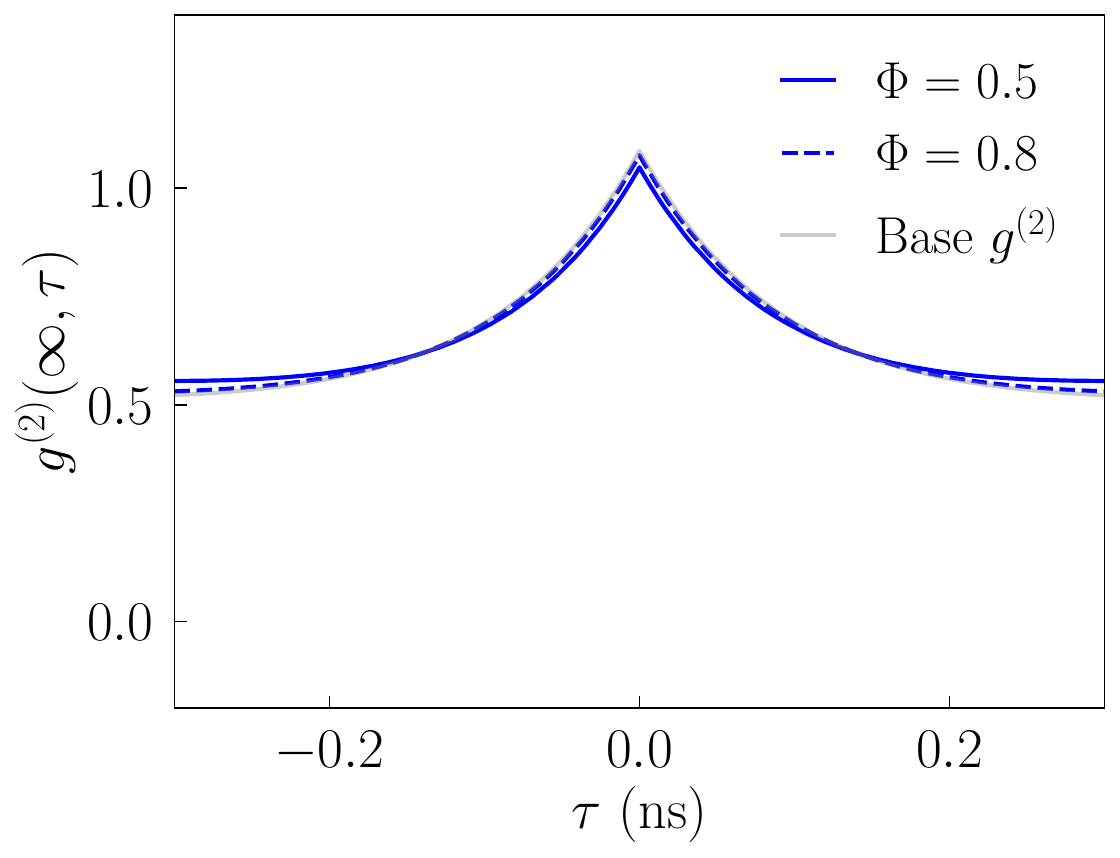}
    
    \includegraphics[width=0.45\textwidth]{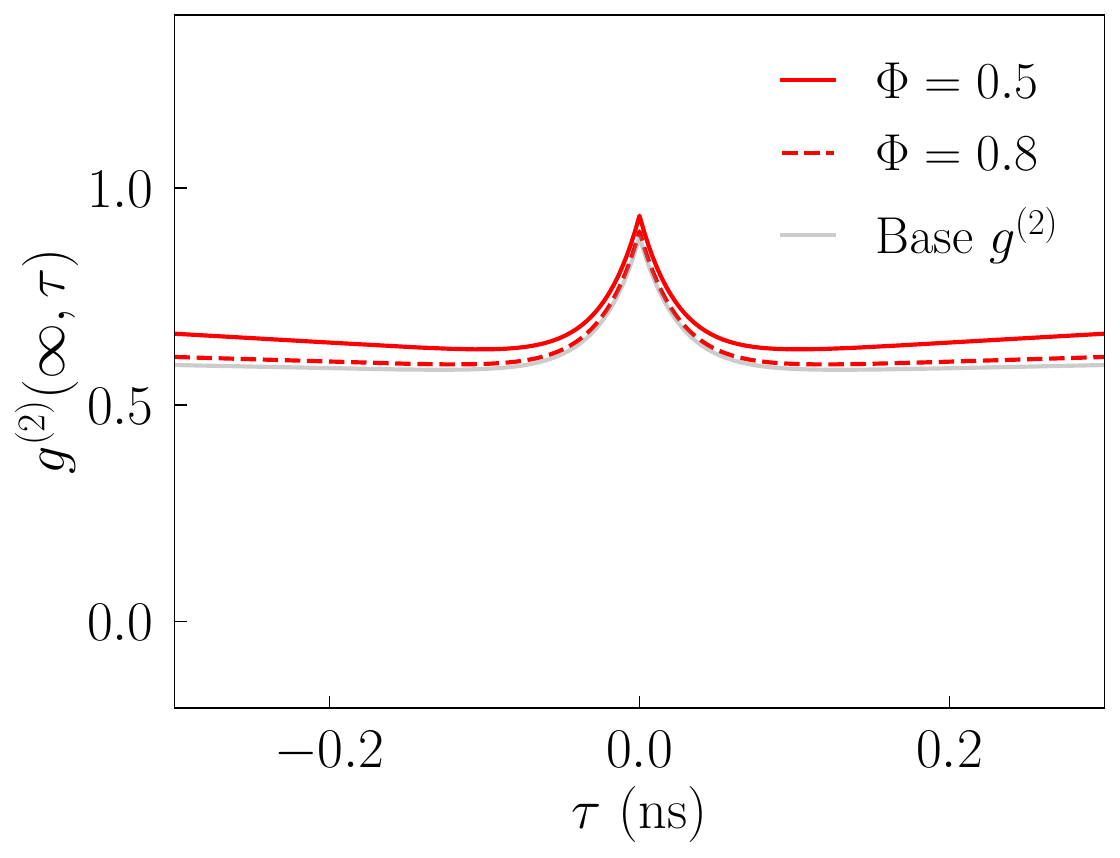}

    \caption{\justifying Effect of site-level non-radiative decay on the photon coincidence, \(g^{(2)}(\infty, \tau)\) for different non-radiative decay rates \(\gamma_{\mathrm{non\mathchar`-rad}}\) leading to different quantum yields, for H-dimers (top row) and J-dimers (bottom row).}
    \label{fig:nonrad}
\end{figure}

Many systems of interest exhibit non-unity quantum yields. For GFP-like parameters (\(\Phi \approx 0.8\)), the non-radiative decay rate satisfies \(\gamma_{\mathrm{non\mathchar`-rad}} \approx 0.25\,\gamma_{\mathrm{rad}}\). We account for this loss channel by introducing Lindblad operators of the form \(\mathcal{L}_i^{(\mathrm{nr})} = \sqrt{\gamma_{\mathrm{non\mathchar`-rad}}}\,\sigma_i^-\), acting locally on each site. As shown in Fig.~\ref{fig:nonrad}, the effect of non-radiative decay on the zero-delay photon coincidence is minimal. Even for \(\Phi = 0.5\), where \(\gamma_{\mathrm{non\mathchar`-rad}} = \gamma_{\mathrm{rad}}\), the anti-dip in \(g^{(2)}(\infty, \tau)\) is reduced by no more than 5\% at zero delay, and typically less than that for GFP-like parameters. This is because, although both radiative and non-radiative processes deplete the symmetric state at similar rates, the symmetric pumping continuously repopulates it. It is worth noting that non-radiative decay slightly decreases the anti-dip height for H-dimers but increases it for J-dimers. This behaviour stems from the different steady-state populations of the bright and dark states in the two configurations (see Sec.~\ref{subsec:ph_coin}), which modifies the relative photon count rates and thus the normalised coincidence signal. At longer delays, minor deviations appear as non-radiative decay alters transient dynamics before the system returns to steady state.

\subsection{Exciton–exciton annihilation}
Exciton–exciton annihilation (EEA) is a key non-radiative process limiting the efficiency of molecular aggregates and organic optoelectronic devices \cite{10.1063/1.4863259}. It arises when multiple excitations are present in close proximity and interact via dipole–dipole coupling, leading to a loss of one excitation and conversion of the other into a higher-energy state. EEA is a two-step process:  (i) adjacent excitons fuse via F\"orster-type coupling into a high-energy doubly excited state, and (ii) this state relaxes non-radiatively to the single-exciton manifold. In the fast-annihilation limit (valid when the relaxation time is much shorter than the radiative lifetime), this can be modelled by direct Lindblad decay channels from the doubly excited state to the single-exciton site basis: \(
\mathcal{L}_{12}^{(\mathrm{ex\mathchar`-an})} 
= \sqrt{\gamma_{\mathrm{ex\mathchar`-an}}}
\ket{e_1 g_2}\bra{e_1 e_2}\) and \(\mathcal{L}_{21}^{(\mathrm{ex\mathchar`-an})} 
= \sqrt{\gamma_{\mathrm{ex\mathchar`-an}}}
\ket{g_1 e_2}\bra{e_1 e_2}
\).
This approach bypasses the need to explicitly include the high-energy intermediate state while capturing its effective dynamics.

\begin{figure}[htbp]
    \centering
    \includegraphics[width=0.45\textwidth]{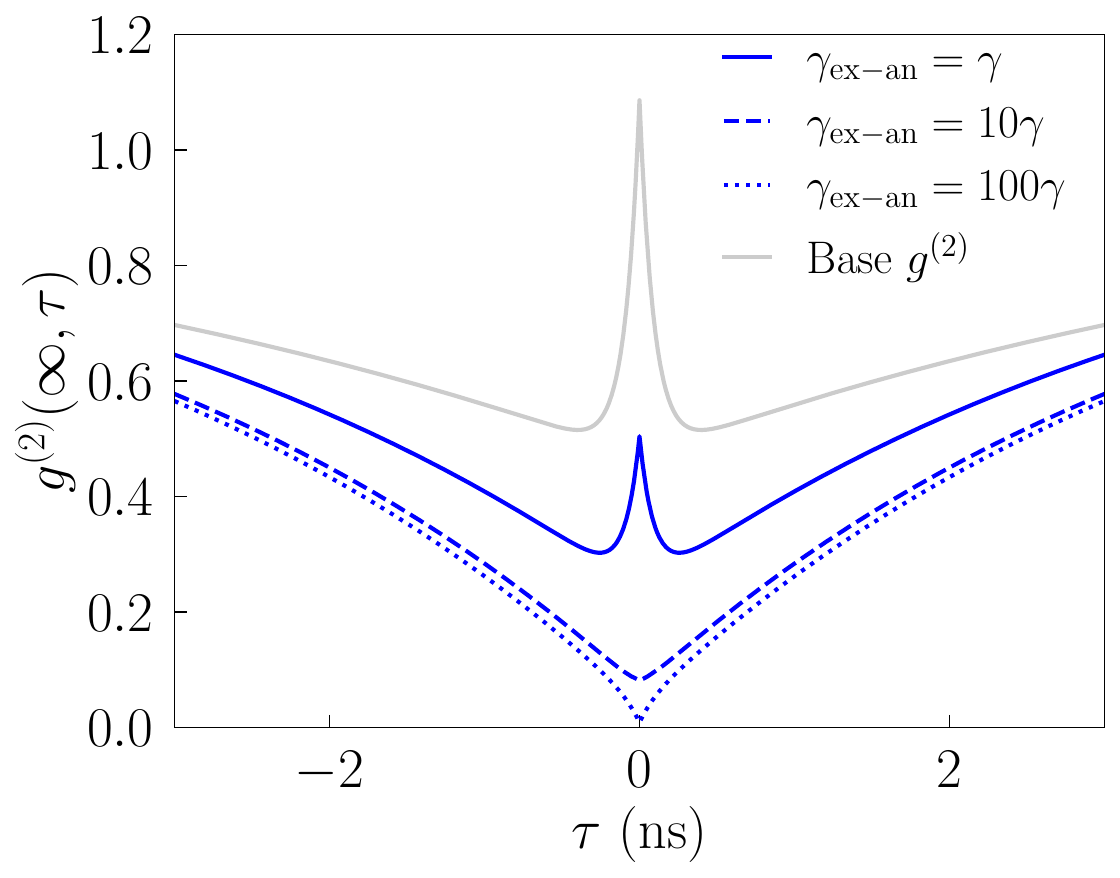}
    
    \includegraphics[width=0.45\textwidth]{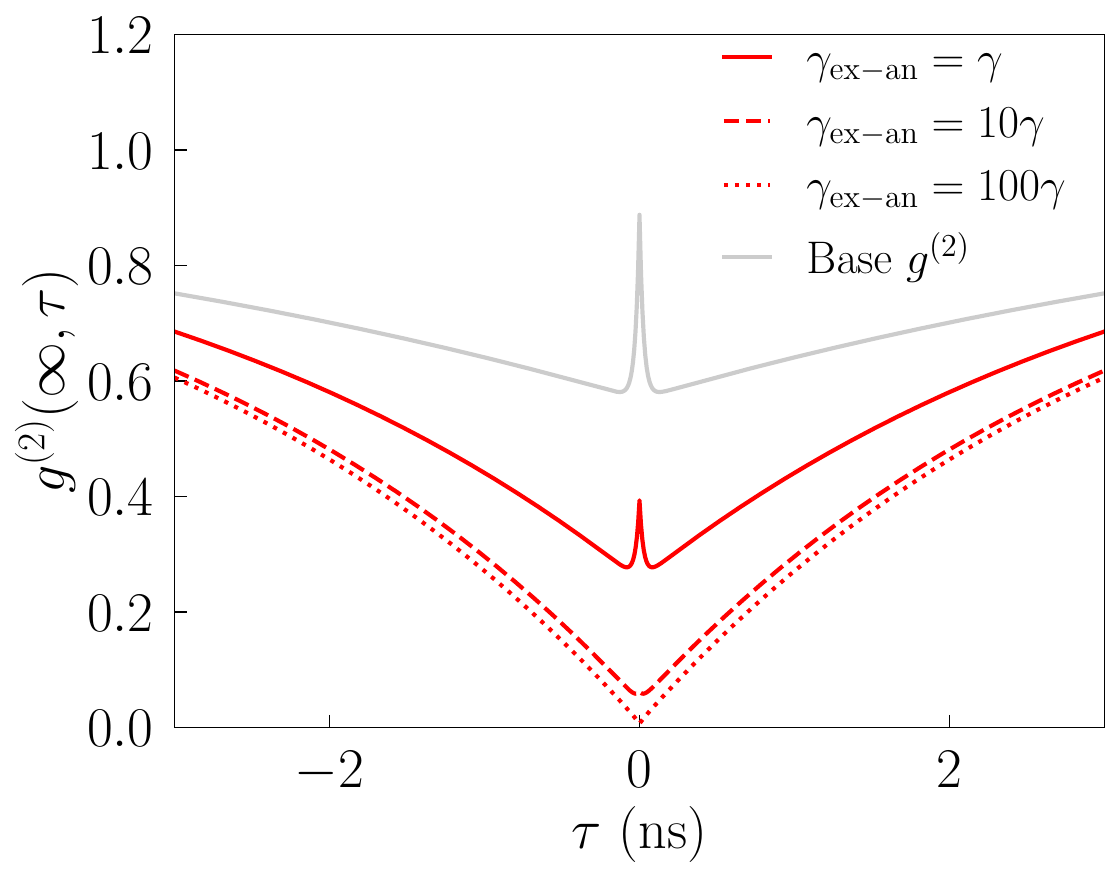}

    \caption{\justifying Impact of exciton–exciton annihilation (EEA) on the photon coincidence, \(g^{(2)}(\infty, \tau)\) for different EEA rates \(\gamma_{\mathrm{ex\mathchar`-an}}\), for H-dimers (top row) and J-dimers (bottom row).}
    \label{fig:exanh}
\end{figure}

Figure~\ref{fig:exanh} shows the resulting zero-delay photon coincidence for H- and J-dimers for different EEA rates in line with experimental estimates \cite{doi:10.1126/sciadv.1600666}. 
As the annihilation rate increases, \(g^{(2)}(\infty,0)\) decreases substantially, 
and for very strong EEA (\(\gamma_{\mathrm{ex\mathchar`-an}} \gg \gamma_{\mathrm{rad}}\)), 
the system approaches single-photon emission behaviour. 
This occurs because biexcitons are rapidly quenched before they can emit the second photon in a cascade. 

Thus, unlike site-level non-radiative decay, exciton–exciton annihilation strongly suppresses two-photon cascades, and \(g^{(2)}(\infty,0)\) is not robust against high annihilation rates.


    \bibliography{references}
    \bibliographystyle{apsrev4-2}
      
\end{document}